\documentclass[letterpaper,twocolumn,10pt]{article}
\usepackage{hyperref}
\PassOptionsToPackage{table}{xcolor}
\usepackage{usenix2019}

\usepackage{booktabs,multirow,multicol}
\usepackage{xspace}
\usepackage{graphicx,xcolor}
\usepackage{cite}
\usepackage[nointegrals]{wasysym}
\usepackage{pifont}
\usepackage{threeparttable}
\usepackage{graphicx}
\newcommand{\cmark}{\ding{51}}%
\newcommand{\xmark}{\ding{55}}%

\usepackage[official]{eurosym}
\usepackage{makecell}
\usepackage{tikz}
\usepackage{amsthm}
\usepackage{enumitem}
\usepackage{url}
\usepackage{amssymb}

\theoremstyle{definition}
\newtheorem{definition}{Definition}[section]

\newcommand{\mypara}[1]{\noindent\textbf{{#1: }}}

\newcommand{\zk}{\text{zk-SNARK}\xspace}

\newcommand{\ZK}{\text{zk-SNARK}\xspace}
\newcommand{\delete}[1]{}
\newcommand{\PCP}{\text{PCP}\xspace}
\newcommand{\IOP}{\text{IOP}\xspace}
\newcommand{\IP}{\text{IP}\xspace}
\newcommand{\PIOP}{\text{PIOP}\xspace}
\newcommand{\PCS}{\text{PCS}\xspace}

\newcommand{\new}[1]{{#1}\xspace}
\newcommand{\news}[1]{{#1}\xspace}

\newcommand{\ZKP}{\text{ZKP}\xspace}
\newcommand{\NIZK}{\text{NIZK}\xspace}
\newcommand{\nlib}{11\xspace}

\newcommand{\recipe}{\textit{master recipe}\xspace}

\newcommand{\lib}[1]{\texttt{#1}\xspace}

\usepackage{color}
\usepackage{tabularx}
\definecolor{darkgreen}{rgb}{0,0.5,0}
\definecolor{darkblue}{rgb}{0,0,0.5}
\definecolor{purple}{rgb}{1,0,1}

\usepackage[most]{tcolorbox}
\usepackage{authblk}

\definecolor{boxcolor}{RGB}{230,240,250}

\newenvironment{takeaway}[1][]
  {
 \begin{tcolorbox}
 [%
    enhanced, 
    breakable,
    boxrule=0.5pt,
    arc=4pt,
    left=2pt,
    right=2pt,
    bottom=2pt,
    top=2pt,
    rounded corners
    ]{}
  \textbf{#1.}
  \small \itshape}
  {
\end{tcolorbox}
}

\begin{document}

\date{}

\title{\Large \bf SoK: Understanding {\zk}s: The Gap Between Research and Practice}

\author{
{\rm Junkai Liang$^{1,*}$, Daqi Hu$^{1,*}$, Pengfei Wu$^{2,*}$, Yunbo Yang$^{3}$, Qingni Shen$^{1,\dag}$, Zhonghai Wu$^{1,\dag}$}\\
$^{1}$Peking University, \quad $^{2}$Singapore Management University, \quad $^{3}$East China Normal University\\
{\tt \{ljknjupku, hudaqi0507\}@gmail.com, pfwu@smu.edu.sg, yyb9882@gmail.com, \\ \tt \{qingnishen, wuzh\}@pku.edu.cn}
} % end author

%\author{
% {\rm Anonymous Author(s)}\\
% \and
% {\rm Junkai Liang}\\
% Peking University
% copy the following lines to add more authors
% \and
% {\rm Name}\\
%Name Institution
%} % end author

\maketitle
\renewcommand{\thefootnote}{}
%-------------------------------------------------------------------------------
\begin{abstract}
\footnote{\rm \textbf{*:} The authors contribute equally to this paper.} \footnote{\rm \textbf{$\dag$:} Corresponding author. This work was supported by the National Key R\&D Program of China under Grant No. 2022YFB2703301, School of Computer Science, Peking University and PKU-OCTA Laboratory for Blockchain and Privacy Computing.}
Zero-knowledge succinct non-interactive argument of knowledge (\zk) serves as a powerful technique for proving the correctness of computations and has attracted significant interest from researchers. Numerous concrete schemes and implementations have been proposed in academia and industry. Unfortunately, the inherent complexity of \zk has created gaps between researchers, developers and users, as they focus differently on this technique. For example, researchers are dedicated to constructing new efficient proving systems with stronger security and new properties. At the same time, developers and users care more about the implementation's toolchains, usability and compatibility. This gap has hindered the development of \zk field.

In this work, we provide a comprehensive study of \zk, from theory to practice, pinpointing gaps and limitations. We first present a \recipe that unifies the main steps in converting a program into a \zk. We then classify existing {\zk}s according to their key techniques. Our classification addresses the main difference in practically valuable properties between existing \zk schemes. We survey over 40 {\zk}s since 2013 and provide a reference table listing their categories and properties. Following the steps in \recipe, we then survey 11 general-purpose popular used libraries. We elaborate on these libraries' usability, compatibility, efficiency and limitations. Since installing and executing these zk-SNARK systems is challenging, we also provide a completely virtual environment in which to run the compiler for each of them. We identify that the proving system is the primary focus in cryptography academia. In contrast, the constraint system presents a bottleneck in industry. To bridge this gap, we offer recommendations and advocate for the open-source community to enhance documentation, standardization and compatibility.

\end{abstract}
\renewcommand{\thefootnote}{\arabic{footnote}}

\section{Introduction}

\new{
	{\textit{Imagine you have a friend who is red-green colour-blind and doubts that red and green are actually distinct colours. You want to prove to your friend that the two colours are indeed different. Our question is: How do you do that without revealing the actual colours of the objects you're using?}
	}
}

\new{The above colour-blind verifier~\cite{colorblind} is a classical problem when thinking about zero-knowledge proof (ZKP) with daily life scenarios. The solution is also easy to understand: 
	You prepare a red ball and a green ball for your friend and ask her to choose one as her favorite. Then she conceals both balls, chooses one ball randomly and asks you to tell if it is her favorite. If red and green are indeed different, you can succeed with probability 1, otherwise, you can only succeed with probability 1/2\footnote{\new{In our simplified question you are not motivated to convince your friend that red and green are the same.}}. Your friend can repeat this process to convince herself that the probability of coincidence is negligible.
}

\begin{figure}[ht]
	\centering
	\includegraphics[height=40mm, width=60mm]{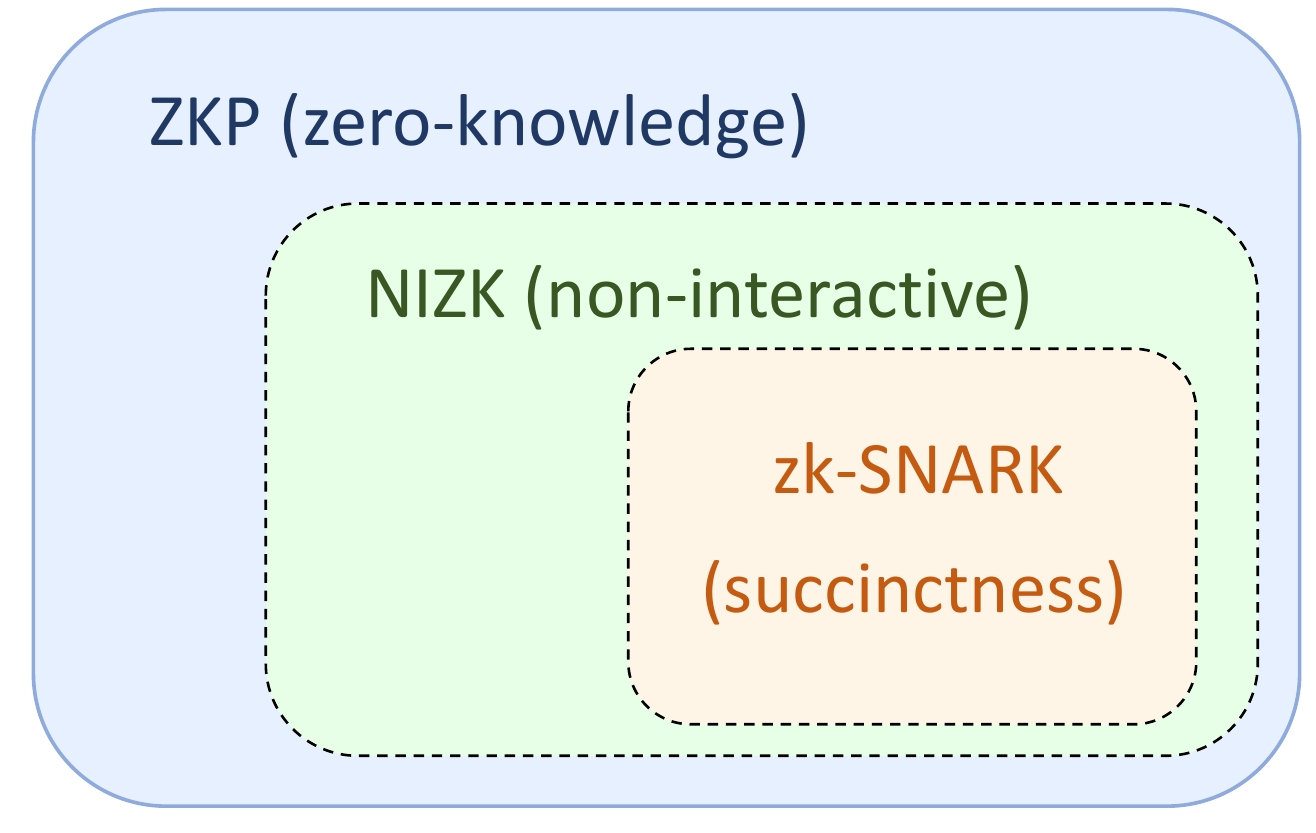}
	\caption{\new{Relations of inclusion for ZKP, NIZK and \ZK.}}
	\label{fig:zkp}
\end{figure}

\new{
	A natural formalism of the above thought experiment yields an interactive form of \ZKP, where there are one or many rounds of interactions between the verifier and the prover~\cite{goldreich1998complexity}, a.k.a. the interactive proof (\IP). \IP is a breakthrough in \ZKP field as it has been used to prove the knowledge of solutions in all problems within non-deterministic polynomial time (NP) space (e.g., 3-colour problem and boolean satisfiability problem), which extends the capability of \ZKP from daily scenarios to computational models~\cite{arora1998proof}. IP is powerful but may need multiple rounds of interaction, which increases the communication burden and is unrealistic for some applications like blockchain or confidential machine learning. Non-interactive zero-knowledge (\NIZK) proof focuses on the protocols where the prover just sends one message (i.e., the proof) to the verifier and the verifier can decide to accept it or not. The main purpose of \NIZK is to solve latency issues caused by interactivity. Luckily, \IP and \NIZK can be bridged through generic transforms, e.g., Fiat-Shamir transform~\cite{fiat1986prove} which allows the prover to generate hash values as if they are random messages given by the verifier. Following the theoretical progress, \IP and \NIZK protocols for the 3-colourability problem and 3-satisfiability have been proposed ~\cite{arora1998proof,arora1998probabilistic}. However, these works suffer from large asymptotic costs and are not practical. To better address real-world scenarios, \NIZK is further required to have succinctness, which means the time and memory used by the prover and verifier are bounded. \NIZK with succinctness, a.k.a. \ZK has been the mainstream of the ZKP research with practical applications. The relations of \ZKP, \NIZK and \ZK are shown in} \autoref{fig:zkp}.

\news{Evolved from \ZKP and \NIZK}, \ZK provides a mechanism for a distrustful party to prove the knowledge of NP relations, where the generated proof reveals nothing about the private witness. This valuable property makes \zk a powerful cryptographic primitive, enabling the verification of computation correctness without exposing private inputs. In the past several years, a surge of groundbreaking scientific achievements has emerged across \zk applications, including but not limited to financial services like blockchain payments~\cite{sasson2014zerocash,bowe2020zexe,bunz2020zether}, smart contract~\cite{wan2022zk,steffen2022zeestar}, and other academic areas like machine learning~\cite{weng2021mystique,liu2021zkcnn}, multiparty computation~\cite{beaver1991secure,ishai2007zero,boyle2019practical} and post-quantum cryptography~\cite{giacomelli2016zkboo,chase2020picnic}. \new{The \ZK also has a promising market outlook. Till today, there are more than 10 widely used blockchains based on \ZK and it has been estimated that only the transaction fee for generating ZK proofs will reach 10 billion by 2030~\cite{zkmarket}. Besides blockchain services, many companies like Axiom \cite{axiom2024}, FedML \cite{fedml2024}, and Giza \cite{giza2024} are cooperating to build ZK ecosystems for privacy-preserving machine learning and other applications. }

\new{
	Despite \ZK having great generality,  succinctness and the potential for wide usage just like encryption and signature algorithms, there are gaps between research and practice that prevent the development of \ZK. 
	Researchers and practitioners have different focuses on three concepts of \ZK: constraint system, proving system, and compiler. Constraint system represents the problems that we want to prove, such as some specific NP relations like the 3-satisfiability. Proving system represents specific cryptographic techniques that generate proof of the relation. Compilers are practical tools that convert a high-level program we want to prove to the constraint system in a mathematical form.
	
	Researchers mainly focus on designing different proving systems for different constraint systems, aiming to achieve special properties. Till today, there are schemes with very practical properties, such as constant proof size, linear prover, post-quantum security, and transparent setup. However, these properties are not integrated into one single scheme and there are trade-offs. To understand these trade-offs, one needs to have substantial knowledge of \ZK mathematical background which is arduous from a practical perspective, preventing a practitioner from choosing an appropriate scheme for her application. Besides, the most time-consuming and error-prone part for practitioners is using the compilers. As reported in \cite{campanelli2017zero,wen2023practical,ozdemir2023bounded,chaliasos2024sok}, programmers struggle to correctly implement their own \ZK applications and there are hundreds of vulnerabilities due to the misunderstanding of the compiler's language.}

We identify a few gaps between academia and industry perspectives in the \zk field: (1) A user requires expert knowledge to choose a scheme, and
% \delete{We find that some libraries provide a mathematical background for their scheme through formulas to persuade their users. However, it is often impossible for users to do a theoretical walk-through.} 
(2) {The importance of the compiler has been underestimated.} To this end, we are interested in the following research questions:
\begin{tcolorbox}
	[%
	enhanced, 
	breakable,
	boxrule=0.5pt,
	arc=4pt,
	left=2pt,
	right=2pt,
	bottom=2pt,
	top=2pt,
	% colback=boxcolor, 
	% colframe=black,
	rounded corners
	% frame hidden,
	% overlay broken = {
		%     \draw[line width=0pt, black, rounded corners]
		%     (frame.north west) rectangle (frame.south east);},
	]{\mypara{RQ1} How to present a unified \emph{master recipe} outlining the design principles and optimizations behind different {\zk}s?
		
		\mypara{RQ2} Can we provide guidelines on selecting {\zk}s in different real-world scenarios?
		
		\mypara{RQ3} From the master recipe and experiments, by scrutinizing prior works, can we provide novel insights for academic researchers and library designers?}
\end{tcolorbox}
\mypara{Our work} To address these questions, we conduct a systematic review of {\zk}s and their libraries. First, we establish a unified master recipe to outline the design principles of mainstream {\zk}s. This recipe includes key steps: compiling a high-level program into a circuit, passing the circuit to a proving system to generate an \IP, and applying a generic transformation to produce the final \zk. Additionally, we explore the main applications of \ZK, such as confidential blockchain, zero-knowledge machine learning, and cryptographic uses.

\new{Using the master recipe, we classify proving systems and trace their evolution in each category.} This helps non-expert users choose suitable \zk schemes. We then evaluate all \nlib state-of-the-art \zk libraries based on performance and usability. By analyzing performance, we recommend best practices for implementing {\zk}s based on different needs. Additionally, we identify common issues in current libraries and advocate for better documentation and standardization.

\news{We emphasize the goal of this paper and its open-source materials aim at four distinct types of readers: (1) researchers who want to move beyond theory to practice by understanding state-of-the-art libraries; (2) developers who want to implement a component as \ZK toolkit; (3) programmers who want to implement their own \ZK applications; and (4) users who want to understand if a certain \ZK application meets their requirements.} 
We believe that our efforts are necessary and can facilitate the practitioners to utilize \zk achievements. 

\mypara{Summary of Contributions} While we are not the first to review this topic, we
position our work as {the first to systematize the research and practice field over the past decade, which tackles the emerging challenges using state-of-the-art libraries.} 
In summary, we have made five main contributions:

\begin{itemize}[noitemsep, topsep=2pt, partopsep=0pt,leftmargin=0.4cm]
	
	\item We establish a unified master recipe showing how a high-level program is converted into a \zk, from the origin to the end. Within the master recipe, we establish a comprehensive overview in \autoref{sec:overview}, considering different circuits, constraint systems, techniques, \new{and applications} used in the practical {\zk}s. 
	
	\item Under the guideline of the master recipe, we further survey more than 40 {\zk}s and provide a comprehensive comparison table for the proving systems. \new{We discuss how the master recipe and the investigation help mitigate the gaps}.
	
	\item We survey all \nlib \zk libraries 
	and make comparisons based on performance and usability. We recommend the best practice implementations and analyze each library's architecture, toolkits and documentation.
	
	\item We provide our well-designed test code examples in docker containers, which we believe will help the development of \zk open source society and users utilize the achievements of \zk field. \new{All our codes and documents are posted on a permanent repository and available at \url{https://doi.org/10.5281/zenodo.14682405}.}
	
	\item Based on comprehensive analyses, we provide key insights and suggestions from 3 perspectives: library selection and programming for non-experts, future directions for researchers, and suggestions for library designers.
\end{itemize}

\mypara{Related Work}
Prior surveys on ZKP fall into two categories. \emph{First}, surveys on \zk constructions and theoretical applications. For example, Feng and Mcllin~\cite{li2014survey} introduce \zk basics and its use for NP computations. Nitulescu~\cite{nitulescu2020zk} focuses on Quadratic Arithmetic Programs (QAP)-based {\zk}s. Li et al.~\cite{LiWei-Han:379} classify {\zk}s by techniques but focus on niche implementations like constraint systems and layered circuits. Others~\cite{morais2019survey,christ2024sok} discuss range proofs and offer practical advice. These works, however, cover only a small portion of {\zk}s and are largely academic. In contrast, our work bridges theory and practice, offering broader insights.
\emph{Second}, surveys on vulnerabilities in practical \zk implementations. Prior works highlight issues in the circuit layer~\cite{wen2023practical,fan2024snarkprobe,isabel2024scalable}, compilation phase~\cite{ozdemir2023bounded}, and application-specific integrity layer~\cite{cerdeira2020sok,zhou2023sok}. Chaliasos et al.~\cite{chaliasos2024sok} summarize these vulnerabilities comprehensively. Our work differs by providing a comprehensive walk-through for \zk practitioners and
focusing on usability, efficiency, compatibility, and library selection, aiming to reduce errors for practitioners unfamiliar with cryptography while emphasizing software security.

\section{Background}
In this section, we focus on the concept of \zk and introduce the definition in \autoref{sec-notion}, as well as the mainstream techniques in \autoref{sec-tech}. In addition, we summarize all abbreviations and their full names in \autoref{abbr}. With these symbols, we discuss the research development of \zk.

\begin{table}[t]
	\centering
	\begin{tabularx}{\linewidth}{p{1.7cm} X}
		\hline
		\textbf{Abbreviation} & \textbf{Full Form} \\
		\hline
		AIR & Arithmetic Intermediate Representation \\
		CRS & Common Reference String \\ 
		DEIP & Doubly Efficient Interactive Proofs \\ 
		(e)DSL & (embedded) Domain-Specific Language \\ 
		FFT & Fast Fourier Transform \\ 
		FRI & {Fast Reed-Solomon IOP of Proximity} \\ 
		HDL & Hardware Description Language\\
		I(O)P & Interactive (Oracle) Proof \\ 
		IPA & Inner Product Argument \\
		ITP & Information-Theoretic Proof \\
		%LPCP & Linear Probabilistically Checkable Proof \\ 
		MPC & Multi-Party Computation \\ 
		NIZK & Non-Interactive Zero-Knowledge \\ 
		NP & Non-deterministic Polynomial Time \\ 
		PL & Programming Language\\
		(L)PCP & (Linear) Probabilistically Checkable Proof \\ 
		PCS & Polynomial Commitment Scheme \\ 
		PIOP & {Polynomial Interactive Oracle Proof} \\ 
		QAP & Quadratic Arithmetic Program \\ 
		QSP & Quadratic Span Program \\ 
		R1CS & Rank-1 Constraint System \\ 
		{STARK} & {Scalable Transparent ARguments of Knowledge} \\ 
		ZKP & Zero-Knowledge Proof \\ 
		ZKML & Zero-Knowledge Machine Learning\\
		{zk-SNARK} & {Zero-Knowledge Succinct Non-Interactive Argument of Knowledge} \\ 
		zk-VM & Zero-Knowledge Virtual Machine \\ 
		\bottomrule
	\end{tabularx}
	\caption{\new{Abbreviations and Corresponding Full Names}}
	\label{abbr}
\end{table}

\subsection{Notions of IP, NIZK and zk-SNARK}
\label{sec-notion}
\new{Here we introduce the formal notions of \IP~\cite{goldwasser2019knowledge}, \NIZK~\cite{groth2010short} and \zk~\cite{gennaro2013quadratic}, which are popular used in the \ZKP field.
	The similarity between these notions is that, for a fixed NP relation $R$, the prover can convince the verifier that for the public input $x$ they know a witness $w$ such that $(x,w)\in R$. The difference is that \IP allows multiple rounds of communication while \NIZK and \zk are non-interactive. Besides, \zk further has efficiency requirements.}
\new{\begin{definition}[\IP]
		Let $R$ be a binary relation induced by a NP language $L$.
		On common input $x$ and prover's input $w$, we denote the interaction between the prover $P$ and the verifier $V$ as $\langle P(w), V\rangle(x)$. A pair $(P,V)$ is called an IP system for $L$ if there exists a negligible function $\epsilon$ such that the following properties hold:
		\begin{itemize}[noitemsep, topsep=2pt, partopsep=0pt,leftmargin=0.4cm]
			\item \textit{Completeness}: If $(x,w)\in R$, then $\Pr[\langle P(w),V\rangle(x)=1]=1$.
			\item \textit{Soundness}: If $(x,w)\notin R$ and for any malicious prover $P^{*}$, we have $\Pr[\langle P^{*}(w),V\rangle(x)=1]<\epsilon(|x|)$. 
		\end{itemize}
\end{definition}}
\begin{definition}[\text{NIZK}]
	{{A NIZK proof consists of three algorithms} $(\textsf{Setup}, \textsf{Prove}, \textsf{Verify})$ that are defined as follows: 
		\begin{itemize}[noitemsep, topsep=2pt, partopsep=0pt,leftmargin=0.4cm]
			\item $\textsf{Setup}(\textsf{pp}) \rightarrow (\textsf{pk},\textsf{vk})$: On input a public parameter $\textsf{pp}$, it outputs a proving and verification key $\textsf{pk}$ and $\textsf{vk}$.
			\item $\textsf{Prove}(\textsf{pk},x,w,R)\rightarrow \pi$: On input $\textsf{pk}$, an instance and witness pair $(x,w)$, and the relation $R$, it outputs a proof $\pi$.
			\item $\textsf{Verify}(\textsf{vk},x,\pi)\rightarrow \{0,1\}$: On input $\textsf{vk},x$, and $\pi$, it outputs 1 or 0 to show if $\pi$ is accepted or not. 
		\end{itemize}
		Besides, a NIZK proof needs to satisfy the following three properties:}
	\begin{itemize}[noitemsep, topsep=2pt, partopsep=0pt,leftmargin=0.4cm]
		\item \textit{Completeness}: Given $(x,w)\in R$, the honest prover results in the verifier outputting 1.
		
		\item \textit{Soundness}: Given $(x,w)\notin R$, a malicious prover interacting with the verifier can only make it output 1 with negligible probability.
		
		\item \textit{Zero knowledge}: Given $(x,w)\in R$, a simulator can produce a view of an honest prover with a possibly malicious verifier that is computationally indistinguishable from an actual execution transcript of the prover with the verifier. Note that the simulator does not get $w$, while the prover gets $w$, so the proof does not contain information of $w$ from the perspective of the verifier.
	\end{itemize}
\end{definition}

\new{A \text{NIZK} proof is termed a \zk if the proof size and verification time are bounded by the size of the statement to be proven:
	
	\begin{itemize}[noitemsep, topsep=2pt, partopsep=0pt,leftmargin=0.4cm]
		\item The proof size is polylogarithmic in the circuit size.
		\item The verification time is polylogarithmic in the circuit size.
	\end{itemize}
}

There are other notions like Scalable Transparent ARguments of Knowledge (\text{STARK})~\cite{ben2019scalable} and Doubly Efficient Interactive Proofs (DEIP)~\cite{wahby2018doubly}, presenting a similar ZKP system like \zk. These notions actually belong to \zk, and the main difference is that they incorporate new properties. For example, \text{STARK} requires a transparent setup, a construction of \zk in the standard model, and post-quantum security; \text {DEIP} requires quasi-linear complexity on the prover side. In this paper, we use \zk to represent the efficient
\text{NIZK} proofs for simplicity.

\subsection{Cryptographic Techniques}
\label{sec-tech}
In this section, we introduce interactive oracle proof (\text{IOP}), which is a generalization of \IP. We also introduce the polynomial commitment scheme (PCS), which can instantiate the oracles in \text{IOP}. We attach great importance to \IOP and \PCS because they help build the structure of the mainstream proving systems. 
We refer to the references~\cite{gabizon2019plonk} for more information, including their concrete constructions.

\begin{definition}[IOP]
	{Let $x$ be a common input known by verifier and prover, $w$ be a witness string only known by prover, and $r(x)\in\mathbb{N}$ be the round complexity on $x$. An IOP system with $r(x)$ rounds asks that for each round, the prover sends a message (which may depend on witness $w$ and prior messages) to the verifier which is given \textit{oracle access}, and the verifier responds with a message to the prover. After interacting with the prover, the output of the verifier is either \texttt{accept} or \texttt{reject}. }
	
	\par Specifically, given $R$ as a binary relation induced by a NP language $L$  and a soundness error $\epsilon \in [0,1]$, we say that a pair of interactive randomized algorithms $(P, V)$ is an IOP system for $L$ with $\epsilon$ if it satisfies the properties below.
	
	\begin{itemize}[noitemsep, topsep=2pt, partopsep=0pt,leftmargin=0.4cm]
		\item \textit{Completeness}: If $(x,w) \in R$, then $\Pr[V(P(x,w),x)= \texttt{accept}]=1$.
		\item \textit{Soundness}: If $(x,w) \notin R$, then for any proof $\pi$, $\Pr[V(\pi,x)= \texttt{accept}]\leq \epsilon$.
	\end{itemize}
\end{definition}

% \begin{definition}[Interactive Oracle Proof (IOP)]
	
	%     For NP relation $R$, soundness error $\epsilon \in [0,1]$, an interactive oracle proof system for $R$ with soundness error $\epsilon$ is a pair of interactive randomized algorithms $(\textsf{P},\textsf{V})$  satisfying the properties below.
	
	%     \begin{itemize}[noitemsep, topsep=2pt, partopsep=0pt,leftmargin=0.4cm]
		%         \item \mypara{Operation} The input of the verifier is $x$, and the input of the prover is $(x,w)$. 
		%         for some witness string $w$. The number of interactive rounds denoted $r(x)$, is called
		%         the round complexity of the system. During a single round, the prover sends
		%         a message (which may depend on witness $w$ and prior messages) to which the verifier
		%         is given \textit{oracle access}, and the verifier responds with a message to the prover. The output of \textsf{V} after interacting with \textsf{P} is either \texttt{accept} or \texttt{reject}.
		
		%         \item \mypara{Completeness} If $(x,w) \in R$, then $\Pr[\textsf{V}(\textsf{P}(x,w),x)= \texttt{accept}]=1$.
		
		%         \item \mypara{Soundness} If $(x,w) \notin R$, then for any proof $\pi$, $\Pr[\textsf{V}(\pi,x)= \texttt{accept}]\leq \epsilon$.
		%     \end{itemize}
	% \end{definition}

As a special case of \text{IOP}, polynomial IOP (\text{PIOP}) denotes a similar interactive process where a proof produces oracles that evaluate polynomials with a degree lower than a given bound.
To ensure privacy, \text{PIOP} is typically instantiated through a PCS, which we define as below.

\begin{definition}[PCS]
	{The PCS allows a prover to commit to a polynomial $f$ %$f\in R[x]$
		and later prove that the committed polynomial was correctly evaluated at a specified point. A PCS consists of four algorithms: $\textsf{Setup}$, $\textsf{Commit}$, $\textsf{Open}$, and $\textsf{VerifyPoly}$.
		\begin{itemize}[noitemsep, topsep=2pt, partopsep=0pt,leftmargin=0.4cm]
			\item $\textsf{Setup}(1^{\kappa})\rightarrow\textsf{ck}$: On input a security parameter $\kappa$, it outputs a commitment key $\textsf{ck}$.
			
			\item $\textsf{Commit}(\textsf{ck},f)\rightarrow\textsf{com}$: On input $\textsf{ck}$ and a polynomial $f$, it outputs a commitment $\textsf{com}$ to $f$.
			
			\item $\textsf{Open}(\textsf{ck},f,\textsf{com},i)\rightarrow {(f(i),\pi)}$: On input $\textsf{ck},f,\textsf{com}$, and a given point $i$, it outputs the evaluation $f(i)$ and a proof $\pi$.
			
			\item $\textsf{VerifyPoly}(\textsf{ck},\textsf{com},i,f(i),\pi)\rightarrow\{0,1\}$: On input $\textsf{ck},\textsf{com},$ $i,f(i)$, and $\pi$, it outputs 1 if $\pi$ is accepted and 0 otherwise.
	\end{itemize}}
\end{definition}

We emphasize \text{PIOP} with \text{PCS} is the mainstream technique in constructing \zk currently. With different instantiations of a \text{PCS}, one can achieve the required properties needed in a \zk (e.g., short proof size, transparency, and post-quantum security). \new{There are also other techniques like the quadratic arithmetic program (\text{QAP}) used to construct a constant-size probabilistically checkable proof (\PCP) as \zk ~\cite{gennaro2013quadratic}. Here, we give a brief introduction to them.}

\new{
	\begin{definition}[PCP]
		\label{PCP}
		Let $R$ be a binary relation induced by a NP language $L$ and $\epsilon\in (0,1)$ be a probability. We say that $R\in PCP(r,q)$ if there is a probabilistic polynomial-time algorithm $V$ for the verifier satisfying the following properties:
		\begin{itemize}[noitemsep, topsep=2pt, partopsep=0pt,leftmargin=0.4cm]
			\item \textit{Efficiency}: After the proof $\pi$ is generated from the witness $w$, $V$ uses at most $r$ random coins and reads at most $q$ bits of $\pi$ to verify it.
			\item \textit{Completeness}: If $(x,w)\in R$, then $\Pr[V(x,\pi)=1]=1$.
			\item \textit{Soundness}: If $x\notin L$, then for all $\pi$, $\Pr[V(x,\pi)=1]<\epsilon$.
		\end{itemize}
	\end{definition}
}

\new{IP, PCP and IOP are all called Information-Theoretic Proof (ITP) which serves as an abstraction of the final \zk scheme. There are two differences among them. First, \IP and \IOP allow interaction without explicitly generating the proof $\pi$, while PCP is non-interactive. Second, \PCP and \IOP use oracles that the verifier can access freely. The oracles serve as a block box to provide additional computation power for the verifier and simplify the protocol design. To help better understand these concepts, we provide a sudoku puzzle example in \autoref{app: ITP}.
}

\new{
	\begin{definition}[QAP]\label{Def-QAP}
		A QAP $Q$ over a field $\mathbb{F}$ involves three sets of $m+1$ polynomials, $L=\{l_{k}(x)\}$, $R=\{r_{k}(x)\}$, $O=\{o_{k}(x)\}$, for $k=\{0,...,m\}$, and a target polynomial $q(x)$. We say that an assignment $(c_1,\ldots,c_m)$ satisfies $Q$ if $q(x)$ divides $p(x)$ (with the quotient denoted as $t(x)$), where
		\begin{equation}
			\begin{aligned}
				p(x)&=L(x)\cdot R(x)-O(x),
			\end{aligned}  
			\label{equ:qap}
		\end{equation}
		$L(x)=l_{0}(x)+\sum_{k=1}^{m}(c_{k}\cdot l_{k}(x))$, $R(x)=r_{0}(x)+\sum_{k=1}^{m}(c_{k}\cdot r_{k}(x))$, and $O(x)=o_{0}(x)+\sum_{k=1}^m(c_{k}\cdot o_{k}(x))$.
	\end{definition}
}

\new{Especially, a circuit with addition and multiplication gates (arithmetic circuit) can be directly represented by QAP by instantiating the polynomials. With this property, QAP has been widely used and abstracted as a constraint system called R1CS. In this paper, we do not distinguish these two concepts.}

\begin{figure*}[t]
	\centering
	\includegraphics[width=\linewidth]{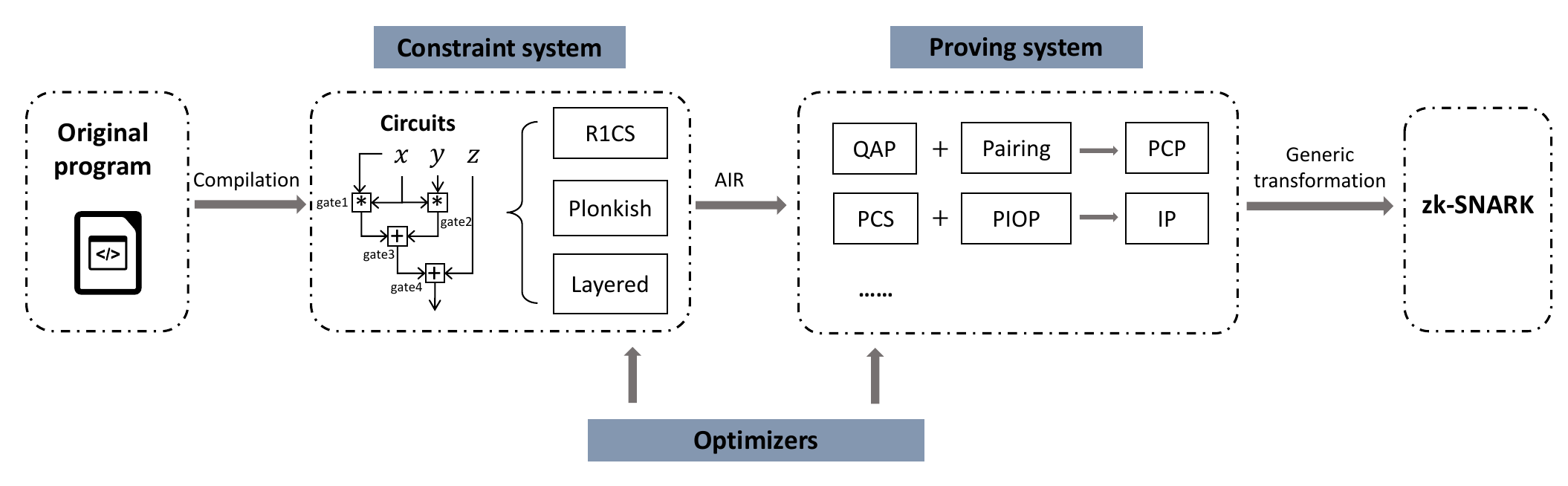}
	\caption{\textbf{The master recipe.} General steps of converting a high-level program to a \zk.}
	\label{fig:overview}
\end{figure*}

\section{Overview}
\label{sec:overview}

In this section, we introduce the \textit{master recipe} of constructing a \zk and discuss the development within each component in \autoref{fig:overview}.
To construct a \zk for general programs, an original program (written in a specific high-level language) is first converted to a circuit form called compilation. Then different constraint systems are utilized to represent the circuit satisfiability problem in mathematical form, a.k.a. Arithmetic Intermediate Representation (\text{AIR}).
Then we need cryptographic protocols to prove the satisfiability of an AIR.
For instance, giving an R1CS, we need an information theoretical protocol to actually prove it. The techniques to instantiate such protocols mainly determine the properties of the final \zk, such as transparency, post-quantum
security and efficiency. They are also our main classification criteria.
Finally, we take a generic transformation to transform the instantiated information theoretical proof into \zk. 
Despite the variations in tools and implementation details, the majority of research topics in \zk fall into our \textit{master recipe}, and we discuss each component in detail as follows.

\mypara{Compiling High-level Programs} \news{Generally, a compiler in \zk implementation compiles a high-level program into AIR that fits a certain constraint system. Currently, the compilers only compile languages that are specific to ZK. These languages are different from the commonly used, general languages like C and Python. Their behaviors are specific to defining a circuit, and the tools and libraries in commonly used languages cannot be recognized by a ZK compiler.}

\mypara{Constraint Systems} With efficient compilers, the high-level program is compiled into the AIR of the circuit, which contains all cryptographic expressions for the relationship between the program's input and output. Generally, a circuit is an abstraction of high-level computation, and a constraint system is a mathematical NP statement that we want to prove. In most cases, these two are similar, and in this paper, we do not distinguish them. \new{Here, we show a classical example where a circuit-like function is transformed to NP language R1CS. Assume we want to prove the computation of $f(w,a,b)=w\cdot(a+b)+(1-w)(a\cdot b)$. If we denote variable $y$ as the output, we can represent the computation by adding variable constraints: $w\cdot(a+b)=y_{1}$, $(1-w)\cdot a=y_{2}$, $b\cdot y_{2}=y_{3}$, $(y_{1}+y_{3})\cdot 1=y$. Following the QAP definition in \autoref{Def-QAP}, the form of R1CS constraint system is: 
	\begin{equation}
		\begin{aligned}
			(l_{0}(x)+\sum_{k=1}^{m}(c_{k}\cdot l_{k}(x)))\cdot(r_{0}(x)+\sum_{k=1}^{m}(c_{k}\cdot r_{k}(x))) \\        
			=(o_{0}(x)+\sum_{k=1}^m(c_{k}\cdot o_{k}(x))).
		\end{aligned}    
	\end{equation}
	
	\noindent Since we totally have 6 variables $w,a,b,y_{1},y_{2},y_{3}$, $m$ is set as 6. Besides, consider that there are 4 constraints. Polynomials $l_{i}, r_{i}$ and $o_{i}$ are evaluated at 4 points and their values should equal the coefficients of the corresponding variable. For instance, let $w$ denotes $c_{1}$, we have $l_{1}(1)=1$ and $l_{1}(2)=-1$, while other points on $l_{1}$ equal $0$ as $w$ does not exist.
}

Common constraint systems include R1CS~\cite{gennaro2013quadratic}, plonk circuit~\cite{gabizon2019plonk} and their variants  such as layered circuits~\cite{xie2019libra,ben2019scalable} and custom plonk~\cite{chen2023hyperplonk}. These constraint systems differ in algebraic structures for high-level computation, making it troublesome for a non-expert developer to understand them completely. \new{For instance, all wire values in plonk circuit are evaluated in one polynomial, while in R1CS the evaluations only encode the existence and coefficients of the variables.}
In most libraries, the languages that define a circuit are related to underlying constraint systems, and developers are required to understand these systems.

\mypara{Proving Systems} 
\new{Proving systems refer to the protocols between the prover and verifier, proving the correctness of a well-defined circuit. {A specific proving system~\cite{gennaro2013quadratic} for the above R1CS example utilizes the bilinear group. The basic idea is that the prover generates group elements $g^{L(x)},g^{R(x)},g^{O(x)}$ and $g^{t(x)}$, then the verifier checks if 
		\begin{equation}\label{equ:proving}
			e(g^{L(x)},g^{R(x)})=e(g^{t(x)},g^{q(x)})\cdot e(g^{O(x)},g),
		\end{equation}
		where $L(x),R(x),O(x),q(x),t(x)$ are defined in \autoref{Def-QAP}, $e$ is bilinear mapping function, and $g$ is the generator of the group.} The advantage of such a proving system is that the proof only consists of a few group elements.}

The proving system is the core component in a \zk and has been widely studied in research. 
A main consideration in choosing proving systems is the desired properties, such as scalability, transparency, post-quantum security and universal setup. Currently, practical {\zk}s with constant proof size and fast verifier are based on QAP techniques~\cite{gennaro2013quadratic,groth2016size} or pairing \PCS~\cite{gabizon2019plonk,chiesa2020marlin}. Those {\zk}s require a trust setup. To eliminate the trust setup, there are {\zk}s utilizing \PCS based on discrete logarithm problem~\cite{bunz2018bulletproofs,bunz2020transparent,halo2book} or hash function with code theory~\cite{chiesa2020fractal,ben2019aurora}. The above schemes all have a slow prover, which is quasi-linear. To achieve a fast prover with linear time, several works~\cite{chen2023hyperplonk,setty2020spartan,golovnev2023brakedown,xie2022orion} design multilinear IOP and multilinear \PCS. However, these approaches utilize more rounds of communication, which significantly increases the proof size. Due to the complicated categories of {\zk}s, it requires expert knowledge of the underlying construction of {\zk}s to choose an appropriate scheme for a particular application. In \autoref{sec:4} we solve this problem by providing a comprehensive classification of existing proving systems.

\mypara{Optimizers}Nowadays, \text{PIOP}-based {\zk}s have achieved the optimized asymptotic complexity for general circuits by introducing linear provers, sublinear proof size and sublinear verifiers. However, the efficiency in specific circumstances can still be improved. For example, recursive~\cite{halo2,plonky2,bunz2020recursive} or aggregate proof~\cite{bunz2018bulletproofs,chung2022bulletproofs+} shrinks the proof size where the verifier needs to verify a sequence of computations. Elastic proof~\cite{bootle2022gemini} and parallel proof~\cite{ephraim2020sparks} allow the prover to adjust the memory and time when proving dynamically. Lookup tables~\cite{campanelli2024lookup} specify the range of the witness to shrink the size of the generating circuit.
It is also possible to improve the performance of modern CPU architecture and specific schemes by optimizing elliptic curve operations~\cite{el2022families}.

\new{
	\mypara{Applications}We can use a general purpose \zk in various applications and prove different computations: (1) In the confidential blockchain, \zk can be utilized to prove a transaction is valid (e.g., if the sender has sufficient funds, the transaction is properly signed and the value is in a certain range) without revealing the details of the transaction to the public, which solves the privacy problem in Bitcoin. Existing blockchain applications include zcash~\cite{sasson2014zerocash}, Ethereum~\cite{wood2014ethereum}, zkSync~\cite{zksync}, and Aztec~\cite{aztecprotocol}, etc. (2) In zero-knowledge machine learning (ZKML), \zk can be used to verify the correctness of training process without revealing the underlying data. This allows the prover to train a model in a verifiable way without sharing her local datasets. Existing ZKML applications focus on generating the proof for decision trees~\cite{zhang2020zero}, federated learning~\cite{duan2024verifiable}, and convolutional neural networks~\cite{liu2021zkcnn}, etc. (3) In cryptography, \zk has been employed to build post-quantum signatures~\cite{chase2020picnic}, verifiable differential privacy mechanisms~\cite{biswas2023interactive}, and oblivious transfer~\cite{hazay2010efficient}, etc. 
}

% \begin{takeaway}[Takeaways]
	%     \textbf{\new{Determine the scope of the work --}}
	%     \new{With the master recipe, a practitioner can better determine the scope of their work, position their problems and understand how the pieces work together as a \zk. }
	
	% \end{takeaway}

\begin{takeaway}[Takeaways]
	\textbf{\new{Determine the scope of the open problems --}}
	\new{With the master recipe, a practitioner can better determine the scope of their work, position their problems and understand how the pieces work together as a \zk. For instance: (1) The latest works which reduce prover time include developing more efficient proof systems, improving circuit compilers and leveraging hardware acceleration (optimizer). (2) In \cite{groth2016size}, a theoretical problem is proposed if three elements are the optimized proof size for \zk. The question is positioned in the proving system and interested readers can focus on its progress without being distracted after understanding the functionality of other components.}
	
\end{takeaway}

\section{Classification of Proving Systems}
\label{sec:4}
\begin{table*}
	
	\renewcommand{\arraystretch}{1.2}
	\setlength{\tabcolsep}{3pt}
	\resizebox{\textwidth}{!}{%
		\begin{threeparttable}
			\begin{tabular}{llllllcllllcl}
				\toprule
				\multicolumn{2}{c}{\textbf{Information Theory}} & \multicolumn{2}{c}{\textbf{Methodology}} & \multicolumn{3}{c}{\textbf{Privacy}} & \multicolumn{3}{c}{\textbf{Scability}} & \multicolumn{1}{c}{\textbf{Examples}} & \multicolumn{1}{c}{\textbf{References}} \\ \cmidrule(lr){1-2}\cmidrule(lr){3-4} \cmidrule(lr){5-7} \cmidrule(lr){8-10} \cmidrule(lr){11-11}
				\cmidrule(lr){12-12}
				\makecell[l]{Type} & \makecell[l]{Variants} & \makecell[l]{Constraint\\ System} & Technique& \makecell[l]{Underlying \\ Problem } &   \makecell[l]{Post \\ Quantum } &
				\makecell[l]{Transparent \\ Setup } &
				\makecell[l]{P Time} & \makecell[l]{V Time} &
				\makecell[l]{Proof Size} &
				&  &  \\ \cline{1-12}
				
				{\makecell[l]{PCP}} & \makecell[l]{LPCP} & R1CS & QAP  & \makecell[l]{q-type\\KoE}& \Circle & \xmark & $\mathcal{O}(N\log N)$ & $\mathcal{O}(l)$ & $\mathcal{O}(1)$ & Groth16 & 
				\cite{gennaro2013quadratic,danezis2013pinocchio,groth2016size,groth2018updatable} \\  
				\cline{1-12}
				
				\multirow{7}{*}{\makecell[l]{IP}} & / &  Layered circuits & GKR & hash& \CIRCLE & \cmark &  $\mathcal{O}(N)$ & $\mathcal{O}(d\log N)$ & $\mathcal{O}(d\log N)$ & \makecell[l]{Virgo, Stark} & {\cite{wahby2018doubly,xie2019libra,zhang2020transparent,zhang2021doubly}} \\ 
				\cline{2-12}
				
				& \multirow{4}{*}{\makecell[l]{PIOP}} & \multirow{4}{*}{\makecell[l]{R1CS/ Plonk}} & KZG PCS & pairing& \Circle &\xmark & $\mathcal{O}(N\log N)$ & $\mathcal{O}(l)$ & $\mathcal{O}(1)$ & Plonk, Marlin & \cite{maller2019sonic,chiesa2020marlin,zhang2017vsql}  \\ 
				
				\cline{4-12}
				&  &  &IPA PCS & \makecell[l]{discrete \\log}& \Circle & \cmark & $\mathcal{O}(N)$ & $\mathcal{O}(\log N)$ & $\mathcal{O}(\log N)$ & \makecell[l]{Halo,\\Bulletproof} & \cite{bunz2018bulletproofs,dalek-bulletproofs,chung2022bulletproofs+,eagen2024bulletproofs++} \\
				\cline{4-12}
				& & & FRI PCS& hash & \CIRCLE & \cmark & $\mathcal{O}(N)$ & $\mathcal{O}(\log^{2}N)$  & $\mathcal{O}(\text{polylog} \ N)$ & Aurora, Fractal & \cite{ben2019aurora,chiesa2020fractal,zhang2021doubly}\\
				\cline{2-12}
				& \makecell[l]{Multi-\\PIOP}& R1CS/ Plonk & Multi-PCS & / & \RIGHTcircle & \cmark & $\mathcal{O}(N)$ & $\mathcal{O}(l)$ & $\mathcal{O}(\log N)$& Hyperplonk, Spartan& \cite{chen2023hyperplonk,setty2020spartan,xie2019libra,xie2022orion,golovnev2023brakedown}\\
				\cline{2-12}
				
				& / &  \makecell[l]{Boolean/Arithmetic \\circuits} & MPC & / & \RIGHTcircle & \cmark &  $\mathcal{O}(N)$ & $\mathcal{O}(N)$ & $\mathcal{O}(N)$ & \makecell[l]{Zkboo} & {\cite{giacomelli2016zkboo,chase2017post,katz2018improved}} \\

				\bottomrule
			\end{tabular}%
		\end{threeparttable}
	}
	\caption{\small \textbf{Classification of ZKPs from different perspectives.} Post Quantum: \Circle: not post-quantum secure, \CIRCLE: plausible post-quantum secure, \RIGHTcircle: partial works in the category are post-quantum secure. Scalability: For R1CS, the circuit size $N$ denotes the number of multiplication gates. For plonk circuit, $N$ is the sum of the addition gate and the multiplication gate. For layered circuits, the circuit size $N=dg$, where $d$ and $g$ are the depth and width of the circuit, respectively. In these circuits, $l$ denote the input size.
		The asymptotic complexity in scalability stands for the optimized scheme in the category.}
	\label{tab:class}
\end{table*}

In this section, we discuss proving systems, the core of \zk field. We classify {\zk}s into two categories termed as \text{PCP} and \text{IP} based on the information-theoretic proof. We discuss the techniques used to construct a \zk in each category and summarize the properties essential for both researchers and developers, such as transparency, post-quantum security, universal setup and efficiency. A comprehensive classification table is provided in~\autoref{tab:class}.
% and a comparison table in ~\autoref{tab:info}.

\subsection{PCP-based {\zk}s}
\label{sec:4.1}
\new{Probabilistically checkable proof (PCP, see \autoref{PCP}) allows for the verification of proofs with extremely high probability by checking only a tiny, randomly chosen portion of the proof. This is in stark contrast to traditional proof verification, which requires reading the entire proof.}

Earlier works~\cite{kilian1992note,groth2010short} of PCPs have high asymptotic complexity and do not focus on general computation models. In 2013, Gennaro et al.~\cite{gennaro2013quadratic} proposed the first efficient {\zk} for general circuits utilizing the quadratic span program (a.k.a. QSP, a weak form of QAP) technique. \new{The basic idea of this category is to construct a set of polynomial equations and use pairings to verify these equations. As an example, to check the validity of \autoref{equ:proving}, one needs four group elements $g^{L(x)}$, $g^{R(x)}$, $g^{O(x)}$ and $g^{t(x)}$ ($q(x)$ can be predefined when instantiating QAP). However, more elements are required to make sure these four elements are indeed computed from the linear combinations of the polynomial coefficients. Besides, we also need to ensure that the same coefficients are used in each linear combination, which we call consistency checks. These checks are based the the Knowledge of Exponent (KoE) assumption~\cite{bellare2004knowledge} and the security guarantee for the group operations is q-type assumption, discussed in \cite{gennaro2013quadratic}. 
	% saying that given two group elements $g,g^\alpha$, it is infeasible to find out another two elements $h,h^\alpha$ without knowing an exponent $c$ such that $h=g^c$ and $h^\alpha=(g^\alpha)^c$. 
	Specifically, the consistency check consists of two aspects:
	% where the verifier sends $(g^{l},g^{\alpha l})$ and the prover returns ($g_{1},g_{2}$) to pass the check $g_{1}^{\alpha}=g_{2}$. If $g_{1}$ is not generated by doing exponentiation from $g^{l}$, such check will fail. The consistency checks are as follows: 
	\begin{itemize}[noitemsep, topsep=2pt, partopsep=0pt,leftmargin=0.4cm]
		\item Polynomial consistency check: The prover computes $g^{L(x)}$ and $g^{\alpha L(x)}$, and the verifier checks if $e(g^{L(x)},g^{\alpha})=e(g^{\alpha L(x)},g)$ holds. For all polynomials, the prover also computes group elements for $R(x),O(x),t(x)$ and carries out this check on them.
		\item Variable consistency check: Given random values $\beta_l,\beta_r,\beta_o$ generated by trusted setup, the prover computes $\prod_{i}^{m} (g^{\beta_{l}l_{i}(x)+\beta_{r}r_{i}(x)+\beta_{o}o_{i}(x)})^{c_{i}}$ as part of the proof, denoted as $g^{Z(x)}$. The verifier checks if $e(g^{L(x)},g^{\beta_{l}\gamma})\cdot e(g^{R(x)},g^{\beta_{r}\gamma})\cdot e(g^{O(x)},g^{\beta_{o}\gamma})=e(g^{Z(x)},g^{\gamma})$.
	\end{itemize}
}

\new{To shrink the proof size, Danezis et al. \cite{danezis2013pinocchio} replace $g^{\beta_{l}}$, $g^{\beta_{r}}$ and $g^{\beta_{o}}$ with three basic group elements $g_{l},g_{r},g_{o}$. Such a replacement saves the need for $\gamma$ and eliminates one element from the proof. In 2016, Groth~\cite{groth2016size} integrated the validity check, polynomial and variable consistency checks into one equation using only three pairings. The proof size was further reduced to an optimized three elements.} Following these theoretical advances, practical work has been done on building concrete implementations. Those works focus on designing a compiler for QAP~\cite{danezis2013pinocchio,ben2013snarks,ben2014succinct}. Since Groth16~\cite{groth2016size} is the optimized QAP-based approach in theory, follow-up works further analyze the security properties~\cite{lipmaa2022unified} and apply it to specific applications together with different models, such as multiparty setup~\cite{bowe2017scalable}, universal reference string (URS)~\cite{groth2018updatable} and recursive proof~\cite{ben2017scalable}. 

The proof size in these systems remains constant, and the time for a prover is linear. These attributes are particularly advantageous and have facilitated real-world implementations, such as ZCash~\cite{sasson2014zerocash} and Pinocchio coin~\cite{danezis2013pinocchio}. Nevertheless, a significant limitation of QAP-based systems is the substantial overhead in prover running time and memory consumption, which poses challenges for scaling to large statements. Additionally, each statement necessitates a separate trusted setup.

\subsection{IP-based {\zk}s}
Interactive proof (\text{IP}) is a generalization of \text{PCP} in which the verifier can send random messages to the prover for multiple rounds.
The construction of IP is divided into two steps: (1) construct a proof which models the message sent by the prover as oracles; and (2) instantiate the oracles with well-defined cryptographic techniques. The first part is also known as PIOP where the prover needs to send a commitment of a polynomial. The technique in the second part is PCS which convinces a verifier that evaluations of a polynomial sent by the prover are correct.
IP can eliminate the trust setup, long common reference string (CRS), and slow prover in QAP-based {\zk}s, and it has been a mainstream in the design of state-of-the-art proving systems.

\subsubsection{GKR-based IP for Layered Circuits}
\label{sec:GKR}
Earlier IPs are mainly designed for layered circuits where each gate can only connect to the layer above. Goldwasser-Kalai-Rothblum (GKR) protocol~\cite{goldwasser2015delegating} is designed to prove the satisfiability of such a circuit by a layer-to-layer reduction. 
\new{The basic idea in this category is that for each layer the prover proves that the gate's output is correctly computed from last layer's output. Denote the number of gates in the $i$-th layer as $S_{i}$ and $s_{i}=\log {S_{i}}$, the label of the wire is $a$, the value of wire $a$ in layer $i$ as $V_{i}(a)$, and the wire predict $\textsf{ADD}_{i}(a,b,c)$ and $\textsf{MUL}_{i}(a,b,c)$ (return 1 when $a,b,c$ combine an addition or multiplication gate, respectively). The GKR prover proves for each wire $c$ in each layer $i$, the following equation holds:
	\begin{equation}
		\label{equ:gkr}
		\begin{aligned}
			V_{i+1}(c)=\sum_{a,b\in \{0,1\}^{s_{i}}} (&\textsf{ADD}_{i}(a,b,c)\cdot (V_{i}(a)+V_{i}(b))\\
			+ \ &\textsf{MUL}_{i}(a,b,c)\cdot V_{i}(a)V_{i}(b))
		\end{aligned}
	\end{equation}
}
\new{\noindent The first GKR protocol has cubic complexity prover, which proves \autoref{equ:gkr} by sending commitments of the circuit values $V_{i}(c)$ and their linear combinations. Several follow-up works~\cite{wahby2018doubly,xie2019libra,zhang2020transparent,zhang2021doubly}
	extend the functions $V,\textsf{ADD},\textsf{MUL}$ in \autoref{equ:gkr} to polynomials as if they are defined in a large field and utilize polynomial evaluations to optimize the complexity to quasi-linear.}
The GKR-based approaches are doubly efficient, meaning that they have a quasi-linear prover along with an efficient verifier where the verifier time is linear to the input of the layered circuit. Despite the advancements of the GKR protocol, a significant limitation is that it only works on layered arithmetic circuits. This introduces a significant overhead when padding general circuits to layered circuits using dummy gates. 

\subsubsection{PIOP for General Circuits}
\label{sec:PIOP}
To construct {\zk}s for general circuits such as R1CS and plonkish circuit, a new construction of IP has been proposed. It utilizes a generalized form of IP called PIOP, which models the message sent by the prover as polynomial oracles, which returns polynomial evaluations. To get an IP, the oracles in PIOP must be instantiated with a PCS, which evaluates a polynomial on a specific point with soundness and privacy. We discuss the features of three different constructions of PCS for univariate PIOP and briefly outline the idea of multivariate PIOP.

\mypara{Univariant PIOP} The idea of univariant PIOP is to model the computation in the general circuit as a polynomial and then prove its properties. \new{The prover uses a polynomial $T$ to encode the values in the whole computation trace, such as the inputs and wire values, and a gate polynomial $S$ to encode all the addition and multiplication gates, e.g., $S(a)=0$ if $a$ is an addition gate and $S(a)=1$ represents a multiplication gate. The prover proves the circuit satisfiability by the following equation for any $y$:
	\begin{equation}
		\label{equ:plonk}
		\begin{aligned}
			S(y)[T(y)+T(\omega y)] + (1-S(y))T(y)T(\omega y)=T(\omega^{2}y),
		\end{aligned}
	\end{equation}
	where $\omega$ is a gate offset, $T(y),T(\omega y),T(\omega^{2}y)$ denote the left input, right input and output of gate $y$, respectively. There are various other polynomial relations related to $T$ and $S$ to ensure the circuit is correct such as zero-test, product-test and permutation-test. All the tests are proved by utilizing PCS, where the prover sends the commitment of these polynomials first and then evaluates them on the point given by the verifier with zero knowledge. The soundness and privacy of all the tests are based on underlying PCS which can fall into three categories.}

\noindent \underline{\textit{{PIOP} with pairing.}} 
The polynomial commitment by Kate, Zaverucha and Goldberg (KZG)~\cite{KZG10} has evaluation proofs that consist of only a single bilinear group element, and verifying an evaluation requires only a single
pairing computation. \new{To evaluate $f(u)=v$ on point $u$, the prover constructs $f(x)-v=(x-u)t(x)$ for some polynomial $t(x)$ and computes the proof as $\pi=g^{t(s)}$, where $s$ is a secret value computed in the trust setup. The verification is done through a pairing operation $e(com/g^{v},g)=e(g^{s}/g^{u},\pi)$ ($com$ is the commitment for the polynomial generated in the setup).
	However, this asymptotically optimal performance comes at the cost of a trusted setup that outputs $g^{s}$ and $s$ must be deleted after generation. 
	
	Many efforts have been made to integrate the KZG PCS into {\zk}s. Plonk~\cite{gabizon2019plonk} utilizes the PCS to evaluate \autoref{equ:plonk}, achieving a short proof and fast quasi-linear prover.
	Similar to Plonk's technique, Marlin~\cite{chiesa2020marlin} applies the KZG PCS to instantiate PIOP to prove the satisfiability of R1CS. It achieves better efficiency for certain types of computation that map well to R1CS (addition gates do not contribute to R1CS's complexity). Some other works~\cite{bunz2021proofs,campanelli2021lunar,zhang2024efficient,aranha2022eclipse} add more features to the \zk in this category like updatable setup and accelerators.
}

\noindent\underline{\textit{{PIOP} with inner-product argument (IPA).}} 
To eliminate the trust setup in pairing-based PCS, BulletProof~\cite{bunz2018bulletproofs} instantiates the \text{PIOP} through a new \text{PCS} using IPA-based techniques. \new{The idea of IPA PCS utilizes algebraic tricks. By proving a polynomial $f$ with degree $m$ equals $v$ at point $u$
	(i.e., $f(u)=\sum_{i=0}^{m}c_{i}u^{i}=v$ where $c_{i}$ is the coefficient), the prover folds the polynomial to two parts as $f(u)=f_{L}(u)+u^{m/2}f_{R}(u)$. By first proving the correctness of the folding and then recursively invoking the procedure, the prover is able to get a logarithmic proof and a linear proving and verifying time related to the polynomial degree.}

\new{Following this technique, Hyrax~\cite{wahby2018doubly} represents the coefficients in a matrix achieving $O(\sqrt{m})$ prover complexity as a refinement. Dory~\cite{lee2021dory} improves the verifier time to logarithmic by introducing a linear combination of the polynomial's coefficients. Other works further optimize the performance in this category achieving both logarithmic time in prover and verifier sides~\cite{bunz2020transparent,wang2022flashproofs,lipmaa2020succinct,arun2023dew}. Several works find IPA PCS is suitable for range proofs and have continued to design optimizers such as aggregate proof, recursive proof and updatable proof in blockchain settings~\cite{bowe2019halo,halo2,chung2022bulletproofs+,eagen2024bulletproofs++,attema2020compressed,wang2022flashproofs,daza2020updateable}. As IPA PCS is based on the hardness of the discrete logarithm problem, the resulting schemes are not post-quantum secure. 
}

\noindent\underline{\textit{{PIOP} with code theory.}} To achieve both transparent setup and post-quantum security, Ligero~\cite{ames2017ligero} utilizes the linear code in code theory to construct a PCS. \new{In linear code, an $[n,k,\Delta]$-code has three properties: (1) it can encode an arbitrary message to a codeword; (2) the minimum distance (Hamming) between any two codewords is $\Delta$; and (3) any linear combination of codewords is also a codeword.
	In Ligero, Reed-Solomon code~\cite{wicker1999reed} is used which views the message as a $k-1$ degree polynomial and views the codeword as its evaluations at $n$ fixed points. In PCS, the $m+1$ coefficients of the polynomial are first encoded into $\mathcal{O}(\sqrt{m})$ codewords. Then the prover commits to the codewords using the Merkle tree to enable the existence check of specific codewords. To verify the evaluation $f(u)=v$, the verifier sends a message $(1,u,\ldots,u^{\mathcal{O}(\sqrt{m})})$ requesting the prover to do linear combinations of the codewords using the message as coefficients. The prover checks (1) the result is generated using the codeword committed before (utilizing the Merkle tree); and (2) the result is a codeword in the same class of the encoding codewords.
	As the message is $\mathcal{O}(\sqrt{m})$-length, the prover size and verifier time both have $\mathcal{O}(\sqrt{m})$ complexity. A bottleneck in the prover side is encoding the polynomial requires FFT which has $\mathcal{O}(\sqrt{m})$ complexity.}

\new{Later works generalize the idea of polynomial encoding by dividing the coefficients in the polynomial into multi-dimensions and encoding them into more codewords~\cite{bootle2020linear,bootle2018efficient} to achieve time-space tradeoff. In~\cite{golovnev2023brakedown}, a different code encoding algorithm is used to further accelerate the prover. 
	In Fractal~\cite{chiesa2020fractal} and other subsequent works~\cite{ben2019aurora,szepieniec2022polynomial}, a novel variant called Fast Reed-Solomon IOP of proximity (\text{FRI})~\cite{ben2018fast} is used. FRI treats the polynomial coefficients as a $\mathcal{O}(m)$-sized vector and recursively encodes it by folding it in half each time to achieve logarithmic proof size. By applying all above-mentioned advanced techniques in code theory, existing code PCS can achieve a logarithmic verifier and proof size, a linear prover and post-quantum security}. 

\mypara{Multivariant PIOP} Though efficient PCS can shrink the proof size and reduce the workload of the verifier, the usage of FFT to construct the key polynomial in the univariate PIOP has been a bottleneck on the prover side as it introduces a quasi-linear complexity. To resolve this efficiency issue, several works~\cite{ chen2023hyperplonk,setty2020spartan,xie2019libra,libert2024simulation,xie2022orion,golovnev2023brakedown} aim at multi-variant polynomial evaluation for eliminating FFT. Those works require modifying the PIOP protocol and PCS to a multivariant type and then using the sumcheck protocol for proving. The key polynomial can be constructed using the multilinear extension technique which only needs linear time.

\mypara{MPC-in-the-head} Several works prove the computation by letting the prover simulate the multiparty protocol~\cite{giacomelli2016zkboo,chase2017post,katz2018improved,ghosal2022efficient,baum2020concretely}.
The technique is called "MPC-in-the-head''. Since it incurs great overhead of the proof size and verifier, this kind of {\zk}s has not been widely implemented.

\begin{takeaway}[Takeaways]
	\textbf{\new{Trade-off between efficiency and security--}}\new{Linear PCP achieves constant proof size but at the cost of a trust setup. The {\zk}s in other categories try to mitigate this issue and all incur a sublinear proof size. In PIOP, compared to the usage of KZG PCS and IPA PCS, the code-based PCS incurs a significant constant overhead in proof size and prover time though the asymptotic complexity is similar.}
	
	\par \textbf{\new{Guidelines for choosing an appropriate proving system--}}
	\new{As a summary of this section, \autoref{tab:class} serves as a guideline for practitioners to choose their appropriate proving systems. We address a few important properties: (1) determine whether a trust setup is accepted. If yes, more considerations shall be taken when choosing the trust third party; (2) determine the appropriate scalability. For instance, blockchain applications prefer a fast verifier and small proof size in order to save transaction fee and the schemes in PIOP with pairing PCS category can be a good choice; and (3) determine if post-quantum security is necessary and choose code-based schemes if yes.}
\end{takeaway}

\section{Library Evaluation}
\label{sec:implementation}

\newcommand{\bin}{\textbf{Binary}}
\newcommand{\ct}{\textbf{CT}}
\begin{table*}[ht]
	\centering
	\resizebox{\textwidth}{!}{
		\begin{threeparttable}
			\rowcolors{2}{gray!20}{white}
			\begin{tabular}{l l l l  c c c l c c c c c c}
				\toprule
				Library & Year & Language & Technique& \makecell[l]{Circuit \\Generality} &  Compiler&\makecell[l]{User\\docus} & \makecell[l]{Example \\docus} & \makecell[l]{Example \\code}  & \makecell[l]{Online support} & \makecell[l]{Academic} &Commercial & \makecell[l]{Last \\update} \\
				\midrule
				
				libsnark~\cite{libsnark} & 2014 & C++ & LPCP-QAP & \cmark & eDSL & \CIRCLE & \Circle & \CIRCLE & \LEFTcircle & \xmark& \xmark& 02/2024 \\
				
				bellman~\cite{bellman} & 2017 & Rust & PIOP-IPA & \cmark & \textbackslash & \Circle & \Circle & \Circle & \Circle &\xmark & \cmark & 07/2024\\
				
				libSTARK~\cite{libsTark} & 2018 & C++ & IP-GKR & \xmark & \textbackslash & \CIRCLE & \Circle & \Circle & \Circle &\cmark & \xmark & 12/2018\\
				
				dalek~\cite{dalek-bulletproofs} & 2018 & Rust & PIOP-IPA & \xmark & \textbackslash & \CIRCLE & \CIRCLE & \CIRCLE & \LEFTcircle &\xmark & \xmark & 01/2024\\
				
				libiop~\cite{libiop} & 2019 & C++ & PIOP-FRI & \xmark & \textbackslash & \CIRCLE & \Circle & \Circle & \Circle &\cmark & \xmark & 05/2021\\
				
				snarkjs~\cite{snarkjs} & 2019 & JavaScript & PCP,PIOP & \cmark & DSL & \CIRCLE & \CIRCLE & \CIRCLE & \CIRCLE &\xmark & \cmark & 04/2024\\
				
				Spartan~\cite{spartan} & 2019 & Rust & PIOP & \cmark & eDSL & \CIRCLE & \LEFTcircle & \CIRCLE & \LEFTcircle &\xmark & \xmark & 04/2024\\
				
				gnark~\cite{gnark} & 2022 & Go & PCP,PIOP & \cmark & eDSL & \CIRCLE & \CIRCLE & \CIRCLE & \CIRCLE &\xmark & \cmark & 07/2024\\
				
				arkworks~\cite{arkworks} & 2022 & Rust & PCP,PIOP & \cmark & DSL & \CIRCLE & \Circle & \CIRCLE & \LEFTcircle &\xmark & \xmark & 01/2023\\
				
				halo2~\cite{halo2} & 2022 & Rust & PIOP-IPA & \cmark & eDSL & \CIRCLE & \CIRCLE & \CIRCLE & \CIRCLE &\xmark & \cmark & 02/2024\\
				
				plonky2~\cite{plonky2} & 2023 & Rust & PIOP & \cmark & eDSL & \CIRCLE & \CIRCLE & \CIRCLE & \CIRCLE &\xmark & \cmark & 08/2024\\
				\bottomrule
			\end{tabular}
	\end{threeparttable}}
	\caption{Comparison table of ZKP implementation libraries. In Circuit generality, \cmark: targets general circuit, \xmark: targets specific circuit. 
		In docus, example codes and online support column, \CIRCLE: full support, \LEFTcircle: partial support, \Circle: lack of support.  }
	\label{tab:info}
\end{table*}
We survey 11 general-purpose popular ZK libraries, all of which contain implementations for \zk protocols aforementioned. Our survey follows the steps in~\autoref{fig:overview} where a high-level program is first converted to an intermediate representation, a.k.a. a circuit, specified by a constraint system. Then, the circuit is passed to a proving system, which implements specific \zk techniques to output a proof. We limit our scope to \zk schemes proposed in the last decade with open-source implementations. Note that the industry in this field is rapidly developing, and some popular protocols, such as halo2~\cite{halo2book} and Plonk~\cite{gabizon2019plonk}, do not have peer-reviewed published papers yet. 
\news{We include those libraries as long as they have basic tools for implementing a circuit (e.g., gadget functions or compiler), their proving systems are popular (at least 5 citations in our references), and they are widely used (e.g., in commercial privacy-focused blockchain projects, or open-source project which have more than 200 GitHub stars and forks).}
% We include those libraries as long as they have enough applications, their corresponding published paper has great influence, or they are widely used in practice (many users are shown in GitHub stars). 
In this section, we compare each library from the perspectives of usability and efficiency\footnote{\new{All our codes and documents are available at \url{https://doi.org/10.5281/zenodo.14682405}.}}. 

\subsection{Basic Information}

We first survey basic information about these libraries, including language, technique, circuit generality, compilers and documentation. Our findings are summarized in~\autoref{tab:info}. The language refers to the programming language that implements the library. The techniques fall into four categories, with PIOP-based schemes being the most common. Circuit generality indicates whether a library supports general circuits. In~\autoref{sec:overview}, we classify R1CS and Plonk circuits as general, while layered circuits and range proofs are not. The latter two can be adapted to general circuits but at an efficiency cost.

Compilers refer to tools that convert high-level languages into circuit constraints, which we categorize in \autoref{sec:compiler}. We also identify valuable documentation types: user documentation (installation, usage, and testing) and example documentation (sample code for applications). Some projects offer additional support via GitHub issues or email.

While some libraries target commercial applications like blockchain transactions, others are research-focused. \news{Due to page limits, detailed discussions on basic information, toolkits, and documentation for each library are provided in \autoref{sec:applib}.}

\subsection{Usability Issues}
Note that some of the attributes in~\autoref{tab:info} represent critical challenges in engineering, which we explain below.

%\begin{itemize}[noitemsep, topsep=2pt, partopsep=0pt,leftmargin=0.4cm]
\mypara{Various Languages and Compatibility}Implementations of \zk schemes are limited across programming languages. For example, Plonk~\cite{gabizon2019plonk} is only implemented in Rust, making it challenging to use in applications written in other languages. Developers needing Plonk-based schemes must use Rust, which may not align with their preferences.
\new{Additionally, none of the libraries provide interfaces for compatibility. While components like constraint systems and proving systems can be separated in code, their functions and tools are confined to their respective libraries. For instance, we attempted to use circuits generated in \lib{libsnark} with \lib{libiop}'s proving systems to test Aurora and Fractal, as suggested by \cite{libiop-issue}. However, we faced significant challenges due to incompatible circuit formats, as there are no interface functions or documentation to bridge the gap.}

\mypara{Misuse of Circuits}Current libraries are not all focused on the general circuits. For instance, Bulletproof~\cite{bunz2018bulletproofs} targets range proofs and is not competitive enough compared with other schemes targeting general circuits like R1CS when designing complex applications. However, an appropriate choice requires expert knowledge of constraint systems, which is impractical for programmers. \news{We believe the master recipe in \autoref{sec:overview} and the classification table and explanations in \autoref{sec:4} can help mitigate this problem by enabling a practitioner to choose an appropriate scheme for her application.}

\new{\mypara{Misuse of Curves}The choice and usage of curves in each library are often implicit, leading 
programmers to overlook this critical configuration. However, selecting an inappropriate curve can reduce efficiency or introduce vulnerabilities. For instance, if the computation exceeds the finite field's limits, the system becomes unsafe, yet programmers may remain unaware. A common example is in blockchain range proofs, where programmers must ensure the curve's bit size exceeds the maximum transaction value; otherwise, severe commercial losses can occur. To address this, we documented the curves used in the surveyed libraries and provided guidelines for proper configuration.
}

\mypara{Lack of Compilers}Many libraries lack a compiler to convert high-level code into circuit representations, forcing programmers to manually add constraints. At the circuit level, programmers must handle intricate details like curve operations, loops, and permutations. For example, implementing a hash function like SHA256 requires tens of thousands of constraints, placing a significant burden on the programmer. Additionally, this task demands deep familiarity with both the programming language and the constraint system.

\mypara{Lack of Documentation}Here, we find that in many libraries, example documents are rather limited. For example, arithmetic circuits operate over a finite field whose size must be set in advance, but very few documents tell how to choose the size. The programmer is responsible for avoiding field overflow, which requires preliminary knowledge of complex field operations.
%\end{itemize}

\new{We have taken steps to address or mitigate these issues. For language and compatibility challenges, we created runnable Docker images for our test sample codes, enabling programmers to configure their environments without relying on cross-platform functions. To tackle circuit and curve misuse, we provided comprehensive guidelines in earlier sections and included a detailed discussion of curves in our project. For compiler-related problems, we categorized existing compilers in each library and analyzed their strengths and weaknesses to help programmers understand compiler concepts in the ZK context. Regarding documentation, we developed open-source materials, including a wiki-book documenting all APIs related to our master recipe components and three walk-through tutorials for our sample code in each library.}

\subsection{Compilers}
\label{sec:compiler}
\new{We identify compilers as the bottleneck of \ZK applications for two reasons. Firstly, during the implementation of our test code, most of the codes are for compilers and we have spent most of time debugging compiler-related issues. Secondly, according to~\cite{chaliasos2024sok}, more than 90\% of the vulnerabilities are found at the circuit level due to misunderstanding the compiler's languages. Here we discuss the categorization of existing compilers for practitioners to understand their features and functionality.
}
\subsubsection{Categorization}
\new{Commonly used compilers for zk are categorized into Domain-Specific Languages (DSLs), Embedded Domain-Specific Languages (eDSLs), and Zero-Knowledge Virtual Machines (zk-VMs). The input of DSL is an independent file with syntax tied to circuit constraints, separate from library functions, and its output is a separate file containing circuit information. The input of eDSL combines library functions related to the constraint system, often using \textbf{gadgets} (built-in functions for complex constraints like inner products or loop specifications); gadgets are tools, not compilers, that help build compiler inputs, and the output of eDSL is a data structure for the proving system. The input of zk-VM is opcodes compiled by general-purpose compilers, and its output is circuit information. We discuss the strengths and drawbacks of these compilers as follows. 
}

%\begin{itemize}[noitemsep, topsep=2pt, partopsep=0pt,leftmargin=0.4cm]

\new{\mypara{Domain-specific languages (DSLs)}DSLs are specialized programming languages designed for specific problem domains, offering tailored syntax to efficiently express constraints in arithmetic circuits for \zk. Current DSLs are categorized as hardware description languages (HDLs)~\cite{belles2022circom} or programming languages (PLs)~\cite{chin2021leo,ozdemir2022circ,amin2023lurk,eberhardt2018zokrates}. HDLs describe circuit synthesis directly in wire form, providing elegant syntax but posing challenges for programmers due to their independent wire-based structure and limited data type abstraction, as inputs are represented as signal data structures. In contrast, PLs define constraints in high-level programming languages, supporting various data types and resembling languages like Rust or Python. This makes PLs more accessible to programmers without wire form circuit knowledge, offering the easiest way to define constraints. However, PLs' flexible syntax increases vulnerability risks and introduces efficiency issues. Currently, learning DSLs is challenging due to the lack of standardization, with each DSL having an entirely different syntax.}

\new{\mypara{Embedded Domain-Specific Languages (\text{eDSL}s)}eDSLs for \zk have gained popularity in recent years and are implemented as functions within general-purpose programming languages, making them distinct from traditional compilers in the context of programming languages. In this paper, we generalize the concept of a compiler to include any tool that transforms its input into a circuit definition. eDSLs are designed to describe circuit synthesis, similar to HDLs, but they target wire form circuits while offering greater expressiveness and ease of use by inheriting data structures and programming features from the embedded language. Examples of eDSLs include implementations in Golang~\cite{gnark}, Rust~\cite{halo2,halo2ce,arkworks,plonky2,zksecurity2023noname,bellman}, C\&C++~\cite{libsnark,libiop}, Java~\cite{kosba2018xjsnark}, and TypeScript~\cite{mina2021o1js}. These eDSLs streamline the development of ZK proofs by integrating circuit definition and proof generation into a single file, simplifying code and enabling programmers to leverage existing library functionalities. However, writing code in eDSLs requires developers to explicitly distinguish between in-circuit and out-circuit operations, necessitating expert knowledge of the specific language and library design.}

\new{\mypara{Zero-Knowledge Virtual Machines (\text{zk-VM}s)}zk-VMs target the opcode of the fetch-decode-execute cycle, replicating the computation trace for general programs (typically smart contracts) and generating corresponding ZK proofs. They support various instruction set architectures (ISAs), including Ethereum Virtual Machine~\cite{scroll2023,polygon,era2023}, RISC~\cite{bruestle2023riscZeroZkVM,arun2024jolt}, and custom ISAs~\cite{goldberg2021cairo,zhang2023polynomial,bootle2018arya}. zk-VMs are compatible with existing high-level programming languages and can leverage features of existing compilers, such as gcc. However, despite targeting low-level opcodes, zk-VMs are not fully compatible with top-level applications and often require minor or major program modifications, which can be error-prone and difficult for programmers to manage. Additionally, zk-VMs use a Turing machine computation model instead of circuits, introducing significant overhead. While zk-VMs reduce the burden of writing constraints for programmers, they may suffer from efficiency issues, particularly for large-scale applications.}

\subsubsection{Compatiability}
\new{We assess the compatibility of these compilers according to two properties:}

\new{
	\mypara{Cross-compatibility} This indicates whether the compilation result of a compiler can be utilized by another one. DSL compilers offer moderate cross-compatibility as they separate the constraint system and the proving system. With the standardization of compilation results in the future, the libraries can only focus on providing the proving systems by taking DSL results as inputs. The eDSL compilers have low cross-compatibility as they define the circuit within a programming language which makes it difficult to use their defined circuit in platforms with other languages. Even in the same language, the compilation result may not be compatible because gadget functions are different, as we find in \lib{libiop} \cite{libiop} and \lib{libsnark} \cite{libsnark}. The zk-VMs have low cross-compatibility as they are only designed for some specific high-level programs. 
	
	\mypara{Syntax-compatibility} This indicates whether the input language of a compiler has a similar syntax to another one. Syntax-compatibility is important as it allows a programmer familiar with a language to move to another one without comprehensive studies. Unfortunately, we find even in the same category, the languages of the compiler have a completely different syntax and it will be hard to learn them all. In DSL, HDL is a hardware circuit language while PL is more like a general programming language. In eDSL, the syntax depends on the basic language of the library, ranging from C, C++, Rust, Go and JavaScript. In zk-VM, only opcodes from smart contract languages are well supported and opcodes from other general languages will not pass the compilation.
}

%\end{itemize} 
\begin{takeaway}[Takeaways]
	\textbf{\new{Absence of universal standardization}} -- \new{Current compilers are categorized as DSL, eDSL and zk-VM, and each has pros and cons. We identify two issues related to compatibility.
		Firstly, even in the same category, there are significant differences in the syntax, which makes it difficult to migrate projects and confuse programmers. Secondly, even for the same circuit, the compilation result cannot be used by a proving system in another library, though the compilers are designed separately from the proving system. We thus call for a universal standardization for these compilers including a standard language syntax and the compilation output.} 
\end{takeaway}

% The purpose of code provided by academia and industry is different. The academia focuses on the construction of a ZKP scheme, so their codes are just to verify the correctness of the scheme, and thus, the codes generally lack modularity and have poor readability and usability. Such as the code provided by ~\cite{giacomelli2016zkboo,wahby2018doubly,maller2019sonic} can only be compile and pass the tests provided by the author, with little documentation and no code description. For some other libraries such as ~\cite{xie2019libra,zhang2020transparent}, we tried to compile them according to the steps provided by the author but failed. While the industry focus on the code's modularity and efficiency, they usually have detailed documentation and teaching steps to facilitate user. That is exactly what we recommend coders to do, and it is also a criterion for us to choose libraries. We surveyed xx libraries, and list xx libraries here due to their extensive documentation, detailed usage steps, hot usage, various use cases or their supplements to some academic libraries.

\begin{table*}[t]
	\resizebox{1\textwidth}{!}{%
		\begin{threeparttable}
			
			\begin{tabular}{ll|lllll|lllll|lllll}
				\toprule
				\multirow{2}{*}{Library} & \multirow{2}{*}{Scheme} & \multicolumn{5}{c|}{Cubic expression} &\multicolumn{5}{c|}{Range proof} &\multicolumn{5}{c}{Hash} \\
				
				& &CRS &N &P &V &S &CRS &N &P &V &S &CRS &N & P &V &S \\
				
				\midrule
				\multirow{3}{*}{\textbf{libsnark}} & \small{Groth16} &0.86 &3 &0.008 &0.001 &0.13 &7.56 &39 &0.023 &0.001 &0.13 &4.19k &27.30k &0.92 &0.001 &0.13\\
				&\small{BCTV14} &1.74 &3 &0.013 &0.004 &0.28 &9.63 &39 &0.024 &0.004 &0.28 &6.28k &27.30k &0.97 &0.004 &0.28\\
				&\small{GM17} &2.11 &3 &0.010 &0.002 &0.13 &15.21 &39 &0.035 &0.002 &0.13 &10.30k &27.30k &1.78 &0.002 &0.13\\
				
				\midrule
				\multirow{2}{*}{\textbf{gnark}} & \small{Groth16} &4.65 &3 &0.002 &0.002 &0.56 &17.09 &22 &0.005 &0.003 &0.70 &100.50k &153.00k &0.28 &{0.002} &0.70 \\
				&\small{Plonk} &31.09 &4 &0.010 &0.004 &1.32 &40.11 &90 &0.003 &0.014 &1.44 &78.91k &599.20k &9.55 &0.002 &1.44\\
				
				\midrule
				\multirow{3}{*}{\textbf{snarkjs}} &\small{Groth16} &5.87 &2 &0.78 &0.70 &0.79 &26.22 &33 &0.76 &0.70 &0.79 &33.00k &59.00k &2.19 &0.71 &0.79 \\
				&\small{Plonk} &13.20 &4 &0.83 &0.71 &2.20 &195.14 &100 &0.94 &0.73 &2.20 &100.50k &241.70k &549.95 &0.76 &2.20  \\
				&\small{FFlonk} &19.67 &4 &0.81 &0.72 &2.20 &291.62 &100 &0.97 &0.70 &2.20 &11044.20k &241.70k &556.31 &0.71 &2.20 \\
				
				\midrule
				\multirow{3}{*}{\textbf{libiop}} &\small{Ligero} &\makecell[c]{\textbackslash} &4 &0.04 &0.01 &608.00 &\makecell[c]{\textbackslash} &32 &0.04 &0.02 &608.00 &\makecell[c]{\textbackslash} &\new{27.28k} &\new{2.195} &\new{2.081} &\new{3.01k} \\
				&\small{Aurora} &\makecell[c]{\textbackslash} &4 &0.022 &0.004 &35.40 &\makecell[c]{\textbackslash} &32 &0.026 &0.007 &50.78 &\makecell[c]{\textbackslash} &\new{32.77k} &\new{7.44} &\new{0.41} &\new{125.98}  \\
				&\small{Fractal} &\makecell[c]{\textbackslash} &4 &0.014 &0.007 &54.69 &\makecell[c]{\textbackslash} &32 &0.044 &0.013 &156.25 &\makecell[c]{\textbackslash} &\new{32.77k} &\new{8.83} &\new{0.012} &\new{201.44} \\
				
				\midrule
				\textbf{Spartan} & \small{Spartan} &\makecell[c]{\textbackslash} &4 &0.59 &0.32 &9.67 &\makecell[c]{\textbackslash} &32 &1.07 &0.45 &15.29 &\makecell[c]{\textbackslash} &32.77k &103.20 &2.07 &67.49 \\
				
				\midrule
				\textbf{arkworks} & \small{Groth16} &2.05 &3 &0.036 &0.033 &0.25 &15.48 &33 &0.037 &0.037 &0.25 &22.32k &58.94k &3.40 &0.036 &0.25 \\
				
				%\small{marlin} &? &? &? &? &? &? &? &? &? &? &? &? &? &? &? \\
				%\small{gm17} &? &? &? &? &? &? &? &? &? &? &? &? &? &? &? \\
				\midrule
				\textbf{halo2} & \small{Halo2} &\makecell[c]{\textbackslash} &2 &0.001 &0.001 &11.97 &\makecell[c]{\textbackslash} &33 &0.002 &0.002 &117.04 &\makecell[c]{\textbackslash} &242.65k &4.16 &0.13 &3.97 \\
				
				\midrule
				\textbf{plonky2} & \small{Plonky2} &\makecell[c]{\textbackslash} &2 &15.36 &0.19 &145.33 &\makecell[c]{\textbackslash} &9 &15.42 &0.19 &145.32 &\makecell[c]{\textbackslash} &261.98k &274.96 &0.28 &175.59 \\
				
				\midrule
				\textbf{dalek}& \small{Bulletproofs} 
				&\multicolumn{5}{c|}{\textbackslash}
				&\multicolumn{2}{c}{\textbackslash} &0.008 &0.001 &0.66 &\multicolumn{5}{c}{\textbackslash} \\
				\bottomrule
			\end{tabular}%
		\end{threeparttable}
		% \vspace{-4mm} 
	}
	\caption{{Main results.
			CRS: the size of a common reference string (KB), N: the number of constraints in a circuit, P: running time of generating a proof (s), V: running time of verifying a proof (s), S: the size of a proof (KB). Since some {\zk}s don't need trust setup, they have no CRS and we mark them with `\textbackslash'. Since Dalek-bulletproofs is used to generate range proofs and not for general circuits, we do not evaluate the Cubic expression or Hash on it.}}
	\label{tab:stat}
\end{table*}

\subsection{Experimental Evaluation}
In this section, we benchmark the performance of {\zk}s on three sample programs. \news{These  programs are all} well-designed and popular in real-world applications. \news{All our experiments are conducted on a server equipped with an Intel Xeon Silver 4314 CPU running at 2.40 GHz. The system is powered by 64 GB of RAM and the operating system used is Ubuntu 20.04.6 LTS.} Our results are reported in \autoref{tab:stat} and here we make two comments. 
\par Firstly, we compare the performance results with the asymptotical complexity of each scheme and give interesting findings that optimal theoretical complexity does not always result in better performance. We discuss why this is the case and recommend researchers discuss more suitable applications for their approach.
\par Secondly, we believe the quantitative
results of our sample programs are meaningful as a reference to practical applications for specific proving systems, but we emphasize that the results would not accurately represent the performance abilities of each scheme. The circuits used in each scheme are different in theory or have different implementations in practice. Besides, the security models vary in transparent setup, post-quantum security, universal reference string, etc., and in academic papers, they are only compared to counterparts in the same category. In our evaluation, we aim to show the common characteristics of each scheme and provide an intuitive comparison from an engineering perspective.

\subsubsection{Sample Programs}

We carefully design sample programs to evaluate the efficiency and usability of each library.

\mypara{A Cubic Expression} Our first example is a cubic expression proof that the prover proves that she knows $x$ that satisfies a polynomial $x^{3}+x+5=y$. This example tests the usability of a library and checks whether it is possible to add constraints manually for arbitrary small-size circuits without compilers. It also tests the basic efficiency of implemented schemes on small-sized circuits.

\mypara{Range Proof} Our second example proves that a value $x$ is in a certain range $[0, 2^{32})$. Range proof is a popular application in blockchain because it enables confidential transactions. Some systems like Bulletproof~\cite{bunz2018bulletproofs} are not designed for general circuits but for range proofs. This example compares such schemes with other general-purpose ones.

\mypara{Hash Function} Our third example is SHA256 hash function. The prover proves that she knows a value $x$ such that $y=\text{SHA256}(x)$ and only $y$ is known to the verifier. A SHA2 hash function is inefficient and has more than 30,000 constraints, which is impossible without compilers. For those libraries that only have proving systems, we test random circuits in the same quantity of constraints instead. The hash example tests the efficiency of the proving systems for large constraints. Additionally, it tests different constant systems, e.g., R1CS and Plonk, when representing the same function.

\subsubsection{Experimental Setup}
In this section, we talk about the criteria for choosing schemes for evaluation and evaluation metrics.

\mypara{Inclusion \& Exclusion Criteria} We aim to build a comprehensive benchmark for more \zk schemes both from papers accepted at top Crypto \& security conferences and industry popular projects (due to the long review cycle, several schemes have not yet been published but have various applications). We then made an initial attempt to run each approach, following the instructions in README on their GitHub homepages and applying the frontend and backend programming styles documented in their evaluation settings. We exclude libraries that either (1) are implemented by authors as materials for the paper or (2) fail to compile and with limited documentation or online support. In the end, we evaluate \textbf{twelve} schemes in {9} libraries, with \textbf{three} (Groth16~\cite{groth2016size}, BCTV14~\cite{ben2014succinct}, GM17~\cite{groth2017snarky}) under the category of QAP (\autoref{sec:4.1}), 
\textbf{one} (Ligero~\cite{ames2017ligero}) under GKR interactive proof (\autoref{sec:GKR}),
and \textbf{eight} (Plonk~\cite{gabizon2019plonk}, Aurora~\cite{ben2019aurora}, Spartan~\cite{setty2020spartan}, Bulletproof~\cite{bunz2018bulletproofs}, Halo2~\cite{halo2book}, 
Plonky2~\cite{plonky2},
Fractal~\cite{chiesa2020fractal},
FFlonk~\cite{gabizon2021fflonk}) 
under \text{PIOP} (\autoref{sec:PIOP}). 

\mypara{Evaluation Metrics} As each scheme has different properties and security models, we choose five general criteria, i.e., (1) the size of common reference string, (2) the number of constraints in the circuit, (3) the running time of the prover, (4) the running time of the verifier, and (5) the size of the proof. We assess different schemes with our three basic examples while more complex examples, such as scenarios that need to verify many proofs at once and a sequence of range proofs, are excluded. Some optimizers like recursive~\cite{halo2} or aggregate proof~\cite{eagen2024bulletproofs++} may perform well in these complex scenarios, but testing them is beyond the scope of this work.
\subsubsection{Performance Highlight}

\mypara{Best Practice} For different application scenarios, we recommend the best scheme along with its implementation. Groth16~\cite{groth2016size} is the best practice for applications that need a fast prover, a small proof size, and can tolerate a trust setup. \lib{gnark}~\cite{gnark} implements Groth16 more efficiently in Go, while \lib{snarkjs}~\cite{snarkjs} provides an implementation of Groth16 in Rust with more compatibility (using a DSL compiler).
Plonk~\cite{gabizon2019plonk} is the best practice for applications that need a transparent setup and are not sensitive to the slight increase of the proof size. For the widely used range proof, we recommend \lib{dalek}~\cite{dalek-bulletproofs}, which is designed for range proof, specifically. We also recommend \lib{gnark}~\cite{gnark}, \lib{arkworks}~\cite{arkworks}, \lib{snarkjs}~\cite{snarkjs}, \lib{halo2}~\cite{halo2} for study or research purposes as they have well-formed documents and running a proof in these libraries follows a complete walk-through of our master recipe.

\section{Discussion}
According to our findings, we advocate for documentation, standardization, and designing specific proving systems, which we explain in detail as follows.

\mypara{Documentation} Universally, the biggest obstacle when using \zk libraries is the lack of documentation. The community has dedicated thousands of hours to producing the work presented here, but not enough documentation makes these contributions less accessible. In the context of \zk field, we recommend two kinds of documents. One is the user document, which contains not only the necessary steps to run an example in the library but also the details of gadgets API in eDSL or the language syntax in DSL about how to define a circuit. The lack of documentation about the compiling phase hinders the library from the cryptographic developers. Besides, We find online support valuable when experimenting with these libraries where the issues in Github solved most of our problems. We also find a walk-through of examples provided by the developers of the library is very helpful. We thus advocate for dynamic documentation such as executable codes (as the docker resources we provided) and enough support through mail or Github. 

\mypara{Standardization} We advocate for two types of standardization. One is about the feature in the \zk field. Many of these libraries are designed around a particular feature, e.g., small proof with a trust setup in \lib{libsnark}~\cite{libsnark}, transparent and fast prover in \lib{arkworks}~\cite{arkworks}. The library's documentation about these core features is implicit, and developers need to understand underlying cryptographic techniques to choose an appropriate scheme. The standardization can help developers compare essential features across libraries and also set a more consistent baseline for performance. The other standardization we advocate is for the compiler. The existing libraries use different approaches such as DSL, eDSL and zk-VMs for defining circuits, which makes it difficult to reuse existing tools due to non-standardization of compilers.

\mypara{Specific Proving Systems} During our exploration, we find some libraries are designed for specific tasks, such as \lib{halo2}~\cite{halo2}, \lib{plonky2}~\cite{plonky2} for recursive proof and \lib{dalek} \cite{dalek-bulletproofs} for range proof. It remains an open question if proving systems for specific scenarios will perform better than generic proving systems. Designing such specific proving systems requires cooperation between the theory progress and engineering.  

\section{Conclusion}

In this paper, we systematically summarized research of \zk from theory to practice. We begin by presenting a master recipe for \zk, which outlines the key steps in constructing {\zk}s. We then examined each component in the recipe from both theoretical and engineering perspectives and identified gaps between them. Extensive efforts were made to evaluate different \zk libraries, and based on our findings, we offered recommendations for programmers and developers while providing new insights for future research.

\section*{Acknowledgments}
The authors appreciate \zk engineers Zhiwen Zhang and Yu'ao Zhou for their help and suggestions when preparing our open-source project.

\section{Ethics Statements and Compliance with the Open Science Policy}

\mypara{Ethics Statements} In this paper, all evaluated \zk libraries are open-source and freely available on GitHub or their respective homepages. As such, this research does not involve any ethical concerns, as it does not include activities that could pose harm or risk to individuals or organizations. We hope this work helps bridge the gap between theory and practice, providing valuable insights for researchers and developers working on \zk applications.

\mypara{Open Science Policy} We fully adhere to the principles of the Open Science Policy and are committed to promoting transparency and reproducibility in scientific research. In line with these principles, we ensure that all evaluated \zk libraries are available with their links provided in the references. 
Our artifacts consist of a completely virtual environment (Docker image), a walk-through tutorial for every test code and an API wiki book in which to run the compiler for each system and are available at \url{https://doi.org/10.5281/zenodo.14682405} as per the conference's requirements. 

\bibliographystyle{unsrt}
\bibliography{def,main}

\begin{thebibliography}{100}

\bibitem{colorblind}
How can convince your colour-blind friend that two balls have the same colour.
\newblock [Online], 2022.
\newblock https://cs.stackexchange.com/questions/150548.

\bibitem{goldreich1998complexity}
Oded Goldreich and Johan H{\aa}stad.
\newblock On the complexity of interactive proofs with bounded communication.
\newblock {\em Inf. Process. Lett.}, 67(4):205--214, 1998.

\bibitem{arora1998proof}
Sanjeev Arora, Carsten Lund, Rajeev Motwani, Madhu Sudan, and Mario Szegedy.
\newblock Proof verification and the hardness of approximation problems.
\newblock {\em Journal of the ACM (JACM)}, 45(3):501--555, 1998.

\bibitem{fiat1986prove}
Amos Fiat and Adi Shamir.
\newblock How to prove yourself: Practical solutions to identification and
  signature problems.
\newblock In {\em Conference on the theory and application of cryptographic
  techniques}, pages 186--194, 1986.

\bibitem{arora1998probabilistic}
Sanjeev Arora and Shmuel Safra.
\newblock Probabilistic checking of proofs: A new characterization of np.
\newblock {\em Journal of the ACM (JACM)}, 45(1):70--122, 1998.

\bibitem{sasson2014zerocash}
Eli~Ben Sasson, Alessandro Chiesa, Christina Garman, Matthew Green, Ian Miers,
  Eran Tromer, and Madars Virza.
\newblock Zerocash: Decentralized anonymous payments from bitcoin.
\newblock In {\em IEEE symposium on security and privacy}, pages 459--474,
  2014.

\bibitem{bowe2020zexe}
Sean Bowe, Alessandro Chiesa, Matthew Green, Ian Miers, Pratyush Mishra, and
  Howard Wu.
\newblock Zexe: Enabling decentralized private computation.
\newblock In {\em 2020 IEEE Symposium on Security and Privacy (SP)}, pages
  947--964, 2020.

\bibitem{bunz2020zether}
Benedikt B{\"u}nz, Shashank Agrawal, Mahdi Zamani, and Dan Boneh.
\newblock Zether: Towards privacy in a smart contract world.
\newblock In {\em International Conference on Financial Cryptography and Data
  Security}, pages 423--443, 2020.

\bibitem{wan2022zk}
Zhiguo Wan, Yan Zhou, and Kui Ren.
\newblock Zk-authfeed: Protecting data feed to smart contracts with
  authenticated zero knowledge proof.
\newblock {\em IEEE Transactions on Dependable and Secure Computing},
  20(2):1335--1347, 2022.

\bibitem{steffen2022zeestar}
Samuel Steffen, Benjamin Bichsel, Roger Baumgartner, and Martin Vechev.
\newblock Zeestar: Private smart contracts by homomorphic encryption and
  zero-knowledge proofs.
\newblock In {\em 2022 IEEE Symposium on Security and Privacy (SP)}, pages
  179--197, 2022.

\bibitem{weng2021mystique}
Chenkai Weng, Kang Yang, Xiang Xie, Jonathan Katz, and Xiao Wang.
\newblock Mystique: Efficient conversions for $\{$Zero-Knowledge$\}$ proofs
  with applications to machine learning.
\newblock In {\em 30th USENIX Security Symposium (USENIX Security 21)}, pages
  501--518, 2021.

\bibitem{liu2021zkcnn}
Tianyi Liu, Xiang Xie, and Yupeng Zhang.
\newblock Zkcnn: Zero knowledge proofs for convolutional neural network
  predictions and accuracy.
\newblock In {\em ACM SIGSAC Conference on Computer and Communications
  Security}, pages 2968--2985, 2021.

\bibitem{beaver1991secure}
Donald Beaver.
\newblock Secure multiparty protocols and zero-knowledge proof systems
  tolerating a faulty minority.
\newblock {\em Journal of Cryptology}, 4:75--122, 1991.

\bibitem{ishai2007zero}
Yuval Ishai, Eyal Kushilevitz, Rafail Ostrovsky, and Amit Sahai.
\newblock Zero-knowledge from secure multiparty computation.
\newblock In {\em Proceedings of the thirty-ninth annual ACM symposium on
  Theory of computing}, pages 21--30, 2007.

\bibitem{boyle2019practical}
Elette Boyle, Niv Gilboa, Yuval Ishai, and Ariel Nof.
\newblock Practical fully secure three-party computation via sublinear
  distributed zero-knowledge proofs.
\newblock In {\em ACM SIGSAC Conference on Computer and Communications
  Security}, pages 869--886, 2019.

\bibitem{giacomelli2016zkboo}
Irene Giacomelli, Jesper Madsen, and Claudio Orlandi.
\newblock $\{$ZKBoo$\}$: Faster $\{$Zero-Knowledge$\}$ for boolean circuits.
\newblock In {\em 25th usenix security symposium (usenix security 16)}, pages
  1069--1083, 2016.

\bibitem{chase2020picnic}
Melissa Chase, David Derler, Steven Goldfeder, Jonathan Katz, Vladimir
  Kolesnikov, Claudio Orlandi, Sebastian Ramacher, Christian Rechberger, Daniel
  Slamanig, Xiao Wang, et~al.
\newblock The picnic signature scheme.
\newblock {\em Submission to NIST Post-Quantum Cryptography project}, 2020.

\bibitem{zkmarket}
Zk market prediction for 2030.
\newblock [Online], 2023.
\newblock https://www.aligned.co/post/10-billion-revenue-market-size-by-2030.

\bibitem{axiom2024}
Axiom.
\newblock Axiom, 2024.
\newblock \url{https://www.axiom.xyz/}.

\bibitem{fedml2024}
FedML.
\newblock Fedml, 2024.
\newblock \url{https://fedml.ai/home}.

\bibitem{giza2024}
Giza.
\newblock Giza, 2024.
\newblock \url{https://gizatech.xyz/}.

\bibitem{campanelli2017zero}
Matteo Campanelli, Rosario Gennaro, Steven Goldfeder, and Luca Nizzardo.
\newblock Zero-knowledge contingent payments revisited: Attacks and payments
  for services.
\newblock In {\em ACM SIGSAC Conference on Computer and Communications
  Security}, pages 229--243, 2017.

\bibitem{wen2023practical}
Hongbo Wen, Jon Stephens, Yanju Chen, Kostas Ferles, Shankara Pailoor, Kyle
  Charbonnet, Isil Dillig, and Yu~Feng.
\newblock Practical security analysis of zero-knowledge proof circuits.
\newblock {\em IACR Cryptol. ePrint Arch.}, 2023:190, 2023.

\bibitem{ozdemir2023bounded}
Alex Ozdemir, Riad~S Wahby, Fraser Brown, and Clark Barrett.
\newblock Bounded verification for finite-field-blasting: In a compiler for
  zero knowledge proofs.
\newblock In {\em International Conference on Computer Aided Verification},
  pages 154--175, 2023.

\bibitem{chaliasos2024sok}
Stefanos Chaliasos, Jens Ernstberger, David Theodore, David Wong, Mohammad
  Jahanara, and Benjamin Livshits.
\newblock Sok: What don't we know? understanding security vulnerabilities in
  snarks.
\newblock {\em arXiv preprint arXiv:2402.15293}, 2024.

\bibitem{li2014survey}
Feng Li and Bruce McMillin.
\newblock A survey on zero-knowledge proofs.
\newblock In {\em Advances in computers}, volume~94, pages 25--69. Elsevier,
  2014.

\bibitem{nitulescu2020zk}
Anca Nitulescu.
\newblock zk-snarks: A gentle introduction.
\newblock {\em Ecole Normale Superieure}, 2020.

\bibitem{LiWei-Han:379}
Li~Wei-Han, ZHANG Zong-Yang, ZHOU Zi-Bo, and DENG Yi.
\newblock An overview on succinct non-interactive zero-knowledge proofs.
\newblock {\em Journal of Cryptologic Research}, 9(3):379--447, 2022.

\bibitem{morais2019survey}
Eduardo Morais, Tommy Koens, Cees Van~Wijk, and Aleksei Koren.
\newblock A survey on zero knowledge range proofs and applications.
\newblock {\em SN Applied Sciences}, 1:1--17, 2019.

\bibitem{christ2024sok}
Miranda Christ, Foteini Baldimtsi, Konstantinos~Kryptos Chalkias, Deepak Maram,
  Arnab Roy, and Joy Wang.
\newblock Sok: Zero-knowledge range proofs.
\newblock {\em Cryptology ePrint Archive}, 2024.

\bibitem{fan2024snarkprobe}
Yongming Fan, Yuquan Xu, and Christina Garman.
\newblock Snarkprobe: An automated security analysis framework for zksnark
  implementations.
\newblock In {\em International Conference on Applied Cryptography and Network
  Security}, pages 340--372, 2024.

\bibitem{isabel2024scalable}
Miguel Isabel, Clara Rodr{\'\i}guez-N{\'u}{\~n}ez, and Albert Rubio.
\newblock Scalable verification of zero-knowledge protocols.
\newblock In {\em IEEE Symposium on Security and Privacy (SP)}, pages 133--133,
  2024.

\bibitem{cerdeira2020sok}
David Cerdeira, Nuno Santos, Pedro Fonseca, and Sandro Pinto.
\newblock Sok: Understanding the prevailing security vulnerabilities in
  trustzone-assisted tee systems.
\newblock In {\em IEEE Symposium on Security and Privacy (SP)}, pages
  1416--1432, 2020.

\bibitem{zhou2023sok}
Liyi Zhou, Xihan Xiong, Jens Ernstberger, Stefanos Chaliasos, Zhipeng Wang,
  Ye~Wang, Kaihua Qin, Roger Wattenhofer, Dawn Song, and Arthur Gervais.
\newblock Sok: Decentralized finance (defi) attacks.
\newblock In {\em 2023 IEEE Symposium on Security and Privacy (SP)}, pages
  2444--2461, 2023.

\bibitem{goldwasser2019knowledge}
Shafi Goldwasser, Silvio Micali, and Chales Rackoff.
\newblock The knowledge complexity of interactive proof-systems.
\newblock In {\em Providing sound foundations for cryptography: On the work of
  shafi goldwasser and silvio micali}, pages 203--225. 2019.

\bibitem{groth2010short}
Jens Groth.
\newblock Short pairing-based non-interactive zero-knowledge arguments.
\newblock In {\em International Conference on the Theory and Application of
  Cryptology and Information Security (AsiaCrypt)}, pages 321--340, 2010.

\bibitem{gennaro2013quadratic}
Rosario Gennaro, Craig Gentry, Bryan Parno, and Mariana Raykova.
\newblock Quadratic span programs and succinct nizks without pcps.
\newblock In {\em International Conference on the Theory and Applications of
  Cryptographic Techniques (EUROCRYPT)}, pages 626--645, 2013.

\bibitem{ben2019scalable}
Eli Ben-Sasson, Iddo Bentov, Yinon Horesh, and Michael Riabzev.
\newblock Scalable zero knowledge with no trusted setup.
\newblock In {\em Advances in Cryptology--CRYPTO 2019: 39th Annual
  International Cryptology Conference}, pages 701--732, 2019.

\bibitem{wahby2018doubly}
Riad~S Wahby, Ioanna Tzialla, Abhi Shelat, Justin Thaler, and Michael Walfish.
\newblock Doubly-efficient zksnarks without trusted setup.
\newblock In {\em 2018 IEEE Symposium on Security and Privacy (SP)}, pages
  926--943, 2018.

\bibitem{gabizon2019plonk}
Ariel Gabizon, Zachary~J Williamson, and Oana Ciobotaru.
\newblock Plonk: Permutations over lagrange-bases for oecumenical
  noninteractive arguments of knowledge.
\newblock {\em Cryptology ePrint Archive}, 2019.

\bibitem{xie2019libra}
Tiacheng Xie, Jiaheng Zhang, Yupeng Zhang, Charalampos Papamanthou, and Dawn
  Song.
\newblock Libra: Succinct zero-knowledge proofs with optimal prover
  computation.
\newblock In {\em Advances in Cryptology--CRYPTO 2019: 39th Annual
  International Cryptology Conference}, pages 733--764, 2019.

\bibitem{chen2023hyperplonk}
Binyi Chen, Benedikt B{\"u}nz, Dan Boneh, and Zhenfei Zhang.
\newblock Hyperplonk: Plonk with linear-time prover and high-degree custom
  gates.
\newblock In {\em International Conference on the Theory and Applications of
  Cryptographic Techniques (EUROCRYPT)}, pages 499--530, 2023.

\bibitem{groth2016size}
Jens Groth.
\newblock On the size of pairing-based non-interactive arguments.
\newblock In {\em International Conference on the Theory and Applications of
  Cryptographic Techniques (EUROCRYPT)}, pages 305--326, 2016.

\bibitem{chiesa2020marlin}
Alessandro Chiesa, Yuncong Hu, Mary Maller, Pratyush Mishra, Noah Vesely, and
  Nicholas Ward.
\newblock Marlin: Preprocessing zksnarks with universal and updatable srs.
\newblock In {\em International Conference on the Theory and Applications of
  Cryptographic Techniques (EUROCRYPT)}, pages 738--768, 2020.

\bibitem{bunz2018bulletproofs}
Benedikt B{\"u}nz, Jonathan Bootle, Dan Boneh, Andrew Poelstra, Pieter Wuille,
  and Greg Maxwell.
\newblock Bulletproofs: Short proofs for confidential transactions and more.
\newblock In {\em 2018 IEEE symposium on security and privacy (SP)}, pages
  315--334, 2018.

\bibitem{bunz2020transparent}
Benedikt B{\"u}nz, Ben Fisch, and Alan Szepieniec.
\newblock Transparent snarks from dark compilers.
\newblock In {\em International Conference on the Theory and Applications of
  Cryptographic Techniques (EUROCRYPT)}, pages 677--706, 2020.

\bibitem{halo2book}
halo2 book.
\newblock \url{https://zcash.github.io/halo2/}, 2022.

\bibitem{chiesa2020fractal}
Alessandro Chiesa, Dev Ojha, and Nicholas Spooner.
\newblock Fractal: Post-quantum and transparent recursive proofs from
  holography.
\newblock In {\em International Conference on the Theory and Applications of
  Cryptographic Techniques (EUROCRYPT)}, pages 769--793, 2020.

\bibitem{ben2019aurora}
Eli Ben-Sasson, Alessandro Chiesa, Michael Riabzev, Nicholas Spooner, Madars
  Virza, and Nicholas~P Ward.
\newblock Aurora: Transparent succinct arguments for r1cs.
\newblock In {\em International Conference on the Theory and Applications of
  Cryptographic Techniques (EUROCRYPT)}, pages 103--128, 2019.

\bibitem{setty2020spartan}
Srinath Setty.
\newblock Spartan: Efficient and general-purpose zksnarks without trusted
  setup.
\newblock In {\em Annual International Cryptology Conference}, pages 704--737,
  2020.

\bibitem{golovnev2023brakedown}
Alexander Golovnev, Jonathan Lee, Srinath Setty, Justin Thaler, and Riad~S
  Wahby.
\newblock Brakedown: Linear-time and field-agnostic snarks for r1cs.
\newblock In {\em Annual International Cryptology Conference}, pages 193--226,
  2023.

\bibitem{xie2022orion}
Tiancheng Xie, Yupeng Zhang, and Dawn Song.
\newblock Orion: Zero knowledge proof with linear prover time.
\newblock In {\em Annual International Cryptology Conference}, pages 299--328,
  2022.

\bibitem{halo2}
ZCash.
\newblock halo2, 2023.
\newblock https://github.com/zcash/halo2.

\bibitem{plonky2}
Mir Protocol.
\newblock Plonky2.
\newblock Github \url{https://github.com/mir-protocol/plonky2}, 2023.

\bibitem{bunz2020recursive}
Benedikt B{\"u}nz, Alessandro Chiesa, Pratyush Mishra, and Nicholas Spooner.
\newblock Recursive proof composition from accumulation schemes.
\newblock In {\em Theory of Cryptography (TCC)}, pages 1--18. Springer, 2020.

\bibitem{chung2022bulletproofs+}
Heewon Chung, Kyoohyung Han, Chanyang Ju, Myungsun Kim, and Jae~Hong Seo.
\newblock Bulletproofs+: shorter proofs for a privacy-enhanced distributed
  ledger.
\newblock {\em Ieee Access}, 10:42081--42096, 2022.

\bibitem{bootle2022gemini}
Jonathan Bootle, Alessandro Chiesa, Yuncong Hu, and Michele Orru.
\newblock Gemini: Elastic snarks for diverse environments.
\newblock In {\em International Conference on the Theory and Applications of
  Cryptographic Techniques (EUROCRYPT)}, pages 427--457, 2022.

\bibitem{ephraim2020sparks}
Naomi Ephraim, Cody Freitag, Ilan Komargodski, and Rafael Pass.
\newblock Sparks: succinct parallelizable arguments of knowledge.
\newblock In {\em Annual International Conference on the Theory and
  Applications of Cryptographic Techniques}, pages 707--737, 2020.

\bibitem{campanelli2024lookup}
Matteo Campanelli, Antonio Faonio, Dario Fiore, Tianyu Li, and Helger Lipmaa.
\newblock Lookup arguments: improvements, extensions and applications to
  zero-knowledge decision trees.
\newblock In {\em International Conference on Public-Key Cryptography (PKC)},
  pages 337--369. Springer, 2024.

\bibitem{el2022families}
Youssef El~Housni and Aurore Guillevic.
\newblock Families of snark-friendly 2-chains of elliptic curves.
\newblock In {\em Annual International Conference on the Theory and
  Applications of Cryptographic Techniques}, pages 367--396, 2022.

\bibitem{wood2014ethereum}
Gavin Wood.
\newblock Ethereum: A secure decentralised generalised transaction ledger.
\newblock {\em Ethereum project yellow paper}, 151:1--32, 2014.

\bibitem{zksync}
zksync.
\newblock \url{https://docs.zksync.io/}, 2023.

\bibitem{aztecprotocol}
arkworks contributors.
\newblock Aztec protocol, 2024.
\newblock https://github.com/AztecProtocol.

\bibitem{zhang2020zero}
Jiaheng Zhang, Zhiyong Fang, Yupeng Zhang, and Dawn Song.
\newblock Zero knowledge proofs for decision tree predictions and accuracy.
\newblock In {\em Proceedings of the 2020 ACM SIGSAC Conference on Computer and
  Communications Security}, pages 2039--2053, 2020.

\bibitem{duan2024verifiable}
Haohua Duan, Zedong Peng, Liyao Xiang, Yuncong Hu, and Bo~Li.
\newblock A verifiable and privacy-preserving federated learning training
  framework.
\newblock {\em IEEE Transactions on Dependable and Secure Computing}, 2024.

\bibitem{biswas2023interactive}
Ari Biswas and Graham Cormode.
\newblock Interactive proofs for differentially private counting.
\newblock In {\em Proceedings of the 2023 ACM SIGSAC Conference on Computer and
  Communications Security}, pages 1919--1933, 2023.

\bibitem{hazay2010efficient}
Carmit Hazay and Yehuda Lindell.
\newblock {\em Efficient secure two-party protocols: Techniques and
  constructions}.
\newblock Springer Science \& Business Media, 2010.

\bibitem{danezis2013pinocchio}
George Danezis, Cedric Fournet, Markulf Kohlweiss, and Bryan Parno.
\newblock Pinocchio coin: building zerocoin from a succinct pairing-based proof
  system.
\newblock In {\em Proceedings of the First ACM workshop on Language support for
  privacy-enhancing technologies}, pages 27--30, 2013.

\bibitem{groth2018updatable}
Jens Groth, Markulf Kohlweiss, Mary Maller, Sarah Meiklejohn, and Ian Miers.
\newblock Updatable and universal common reference strings with applications to
  zk-snarks.
\newblock In {\em Annual International Cryptology Conference}, pages 698--728,
  2018.

\bibitem{zhang2020transparent}
Jiaheng Zhang, Tiancheng Xie, Yupeng Zhang, and Dawn Song.
\newblock Transparent polynomial delegation and its applications to zero
  knowledge proof.
\newblock In {\em IEEE Symposium on Security and Privacy (SP)}, pages 859--876,
  2020.

\bibitem{zhang2021doubly}
Jiaheng Zhang, Tianyi Liu, Weijie Wang, Yinuo Zhang, Dawn Song, Xiang Xie, and
  Yupeng Zhang.
\newblock Doubly efficient interactive proofs for general arithmetic circuits
  with linear prover time.
\newblock In {\em ACM SIGSAC Conference on Computer and Communications
  Security}, pages 159--177, 2021.

\bibitem{maller2019sonic}
Mary Maller, Sean Bowe, Markulf Kohlweiss, and Sarah Meiklejohn.
\newblock Sonic: Zero-knowledge snarks from linear-size universal and updatable
  structured reference strings.
\newblock In {\em ACM SIGSAC Conference on Computer and Communications
  Security}, pages 2111--2128, 2019.

\bibitem{zhang2017vsql}
Yupeng Zhang, Daniel Genkin, Jonathan Katz, Dimitrios Papadopoulos, and
  Charalampos Papamanthou.
\newblock vsql: Verifying arbitrary sql queries over dynamic outsourced
  databases.
\newblock In {\em 2017 IEEE Symposium on Security and Privacy (SP)}, pages
  863--880, 2017.

\bibitem{dalek-bulletproofs}
dalek contributors.
\newblock dalek-bulletproof, 2017.
\newblock https://github.com/dale-cryptography/bulletproofs.

\bibitem{eagen2024bulletproofs++}
Liam Eagen, Sanket Kanjalkar, Tim Ruffing, and Jonas Nick.
\newblock Bulletproofs++: next generation confidential transactions via
  reciprocal set membership arguments.
\newblock In {\em International Conference on the Theory and Applications of
  Cryptographic Techniques (EUROCRYPT)}, pages 249--279, 2024.

\bibitem{chase2017post}
Melissa Chase, David Derler, Steven Goldfeder, Claudio Orlandi, Sebastian
  Ramacher, Christian Rechberger, Daniel Slamanig, and Greg Zaverucha.
\newblock Post-quantum zero-knowledge and signatures from symmetric-key
  primitives.
\newblock In {\em ACM SIGSAC Conference on Computer and Communications
  Security}, pages 1825--1842, 2017.

\bibitem{katz2018improved}
Jonathan Katz, Vladimir Kolesnikov, and Xiao Wang.
\newblock Improved non-interactive zero knowledge with applications to
  post-quantum signatures.
\newblock In {\em ACM SIGSAC Conference on Computer and Communications
  Security}, pages 525--537, 2018.

\bibitem{kilian1992note}
Joe Kilian.
\newblock A note on efficient zero-knowledge proofs and arguments.
\newblock In {\em Proceedings of the twenty-fourth annual ACM symposium on
  Theory of computing}, pages 723--732, 1992.

\bibitem{bellare2004knowledge}
Mihir Bellare and Adriana Palacio.
\newblock The knowledge-of-exponent assumptions and 3-round zero-knowledge
  protocols.
\newblock In {\em Annual International Cryptology Conference}, pages 273--289.
  Springer, 2004.

\bibitem{ben2013snarks}
Eli Ben-Sasson, Alessandro Chiesa, Daniel Genkin, Eran Tromer, and Madars
  Virza.
\newblock Snarks for c: Verifying program executions succinctly and in zero
  knowledge.
\newblock In {\em Annual cryptology conference}, pages 90--108, 2013.

\bibitem{ben2014succinct}
Eli Ben-Sasson, Alessandro Chiesa, Eran Tromer, and Madars Virza.
\newblock Succinct $\{$Non-Interactive$\}$ zero knowledge for a von neumann
  architecture.
\newblock In {\em 23rd USENIX Security Symposium (USENIX Security 14)}, pages
  781--796, 2014.

\bibitem{lipmaa2022unified}
Helger Lipmaa.
\newblock A unified framework for non-universal snarks.
\newblock In {\em International Conference on Public-Key Cryptography (PKC)},
  pages 553--583. Springer, 2022.

\bibitem{bowe2017scalable}
Sean Bowe, Ariel Gabizon, and Ian Miers.
\newblock Scalable multi-party computation for zk-snark parameters in the
  random beacon model.
\newblock {\em Cryptology ePrint Archive}, 2017.

\bibitem{ben2017scalable}
Eli Ben-Sasson, Alessandro Chiesa, Eran Tromer, and Madars Virza.
\newblock Scalable zero knowledge via cycles of elliptic curves.
\newblock {\em Algorithmica}, 79:1102--1160, 2017.

\bibitem{goldwasser2015delegating}
Shafi Goldwasser, Yael~Tauman Kalai, and Guy~N Rothblum.
\newblock Delegating computation: interactive proofs for muggles.
\newblock {\em Journal of the ACM (JACM)}, 62(4):1--64, 2008.

\bibitem{KZG10}
Aniket Kate, Gregory~M. Zaverucha, and Ian Goldberg.
\newblock Constant-size commitments to polynomials and their applications.
\newblock In {\em Advances in Cryptology - ASIACRYPT 2010}, pages 177--194,
  Berlin, Heidelberg, 2010.

\bibitem{bunz2021proofs}
Benedikt B{\"u}nz, Mary Maller, Pratyush Mishra, Nirvan Tyagi, and Psi Vesely.
\newblock Proofs for inner pairing products and applications.
\newblock In {\em International Conference on the Theory and Application of
  Cryptology and Information Security (AsiaCrypt)}, pages 65--97, 2021.

\bibitem{campanelli2021lunar}
Matteo Campanelli, Antonio Faonio, Dario Fiore, Ana{\"\i}s Querol, and
  Hadri{\'a}n Rodr{\'\i}guez.
\newblock Lunar: a toolbox for more efficient universal and updatable zksnarks
  and commit-and-prove extensions.
\newblock In {\em International Conference on the Theory and Application of
  Cryptology and Information Security (AsiaCrypt)}, pages 3--33. Springer,
  2021.

\bibitem{zhang2024efficient}
Yuncong Zhang, Shi-Feng Sun, and Dawu Gu.
\newblock Efficient kzg-based univariate sum-check and lookup argument.
\newblock In {\em International Conference on Public-Key Cryptography (PKC)},
  pages 400--425. Springer, 2024.

\bibitem{aranha2022eclipse}
Diego~F Aranha, Emil~Madsen Bennedsen, Matteo Campanelli, Chaya Ganesh, Claudio
  Orlandi, and Akira Takahashi.
\newblock Eclipse: enhanced compiling method for pedersen-committed zksnark
  engines.
\newblock In {\em International Conference on Public-Key Cryptography (PKC)},
  pages 584--614. Springer, 2022.

\bibitem{lee2021dory}
Jonathan Lee.
\newblock Dory: Efficient, transparent arguments for generalised inner products
  and polynomial commitments.
\newblock In {\em Theory of Cryptography (TCC)}, pages 1--34, 2021.

\bibitem{wang2022flashproofs}
Nan Wang and Sid Chi-Kin Chau.
\newblock Flashproofs: Efficient zero-knowledge arguments of range and
  polynomial evaluation with transparent setup.
\newblock In {\em International Conference on the Theory and Application of
  Cryptology and Information Security (AsiaCrypt)}, pages 219--248, 2022.

\bibitem{lipmaa2020succinct}
Helger Lipmaa and Kateryna Pavlyk.
\newblock Succinct functional commitment for a large class of arithmetic
  circuits.
\newblock In {\em International Conference on the Theory and Application of
  Cryptology and Information Security (AsiaCrypt)}, pages 686--716. Springer,
  2020.

\bibitem{arun2023dew}
Arasu Arun, Chaya Ganesh, Satya Lokam, Tushar Mopuri, and Sriram Sridhar.
\newblock Dew: a transparent constant-sized polynomial commitment scheme.
\newblock In {\em International Conference on Public-Key Cryptography (PKC)},
  pages 542--571. Springer, 2023.

\bibitem{bowe2019halo}
Sean Bowe, Jack Grigg, and Daira Hopwood.
\newblock Halo: Recursive proof composition without a trusted setup. iacr
  cryptol. eprint arch.(2019), 1021.
\newblock {\em URL: https://eprint. iacr. org/2019/1021}, 2019.

\bibitem{attema2020compressed}
Thomas Attema and Ronald Cramer.
\newblock Compressed-protocol theory and practical application to plug \& play
  secure algorithmics.
\newblock In {\em Annual International Cryptology Conference}, pages 513--543,
  2020.

\bibitem{daza2020updateable}
Vanesa Daza, Carla R{\`a}fols, and Alexandros Zacharakis.
\newblock Updateable inner product argument with logarithmic verifier and
  applications.
\newblock In {\em International Conference on Public-Key Cryptography (PKC)},
  pages 527--557, 2020.

\bibitem{ames2017ligero}
Scott Ames, Carmit Hazay, Yuval Ishai, and Muthuramakrishnan
  Venkitasubramaniam.
\newblock Ligero: Lightweight sublinear arguments without a trusted setup.
\newblock In {\em ACM SIGSAC Conference on Computer and Communications
  Security}, pages 2087--2104, 2017.

\bibitem{wicker1999reed}
Stephen~B Wicker and Vijay~K Bhargava.
\newblock {\em Reed-Solomon codes and their applications}.
\newblock John Wiley \& Sons, 1999.

\bibitem{bootle2020linear}
Jonathan Bootle, Alessandro Chiesa, and Jens Groth.
\newblock Linear-time arguments with sublinear verification from tensor codes.
\newblock In {\em Theory of Cryptography (TCC)}, pages 19--46. Springer, 2020.

\bibitem{bootle2018efficient}
Jonathan Bootle and Jens Groth.
\newblock Efficient batch zero-knowledge arguments for low degree polynomials.
\newblock In {\em IACR International Workshop on Public Key Cryptography},
  pages 561--588. Springer, 2018.

\bibitem{szepieniec2022polynomial}
Alan Szepieniec and Yuncong Zhang.
\newblock Polynomial iops for linear algebra relations.
\newblock In {\em International Conference on Public-Key Cryptography (PKC)},
  pages 523--552. Springer, 2022.

\bibitem{ben2018fast}
Eli Ben-Sasson, Iddo Bentov, Yinon Horesh, and Michael Riabzev.
\newblock Fast reed-solomon interactive oracle proofs of proximity.
\newblock In {\em 45th international colloquium on automata, languages, and
  programming (icalp 2018)}, 2018.

\bibitem{libert2024simulation}
Beno{\^\i}t Libert.
\newblock Simulation-extractable kzg polynomial commitments and applications to
  hyperplonk.
\newblock In {\em International Conference on Public-Key Cryptography (PKC)},
  pages 68--98. Springer, 2024.

\bibitem{ghosal2022efficient}
Riddhi Ghosal, Paul Lou, and Amit Sahai.
\newblock Efficient nizks from lwe via polynomial reconstruction and “mpc in
  the head”.
\newblock In {\em International Conference on the Theory and Application of
  Cryptology and Information Security (AsiaCrypt)}, pages 496--521. Springer,
  2022.

\bibitem{baum2020concretely}
Carsten Baum and Ariel Nof.
\newblock Concretely-efficient zero-knowledge arguments for arithmetic circuits
  and their application to lattice-based cryptography.
\newblock In {\em International Conference on Public-Key Cryptography (PKC)},
  pages 495--526. Springer, 2020.

\bibitem{libsnark}
libsnark contributors.
\newblock libsnark, 2014.
\newblock https://github.com/scipr-lab/libsnark.

\bibitem{bellman}
bellman contributors.
\newblock zkcrypto/bellman, 2017.
\newblock https://github.com/zkcrypto/bellman.

\bibitem{libsTark}
libSTARK contributors.
\newblock elibensasson/libstark, 2018.
\newblock https://github.com/elibensasson/libSTARK.

\bibitem{libiop}
libiop contributors.
\newblock libiop, 2019.
\newblock https://github.com/scipr-lab/libiop.

\bibitem{snarkjs}
snarkjs contributors.
\newblock snarkjs, 2020.
\newblock https://github.com/iden3/snarkjs.

\bibitem{spartan}
spartan contributors.
\newblock microsoft/spartan, 2020.
\newblock https://github.com/microsoft/Spartan.

\bibitem{gnark}
gnark contributors.
\newblock gnark, 2022.
\newblock https://github.com/Consensys/gnark.

\bibitem{arkworks}
arkworks contributors.
\newblock \texttt{arkworks} zksnark ecosystem, 2022.
\newblock https://arkworks.rs.

\bibitem{libiop-issue}
How can convince your colour-blind friend that two balls have the same colour.
\newblock [Online], 2019.
\newblock https://github.com/scipr-lab/libiop/issues/2.

\bibitem{belles2022circom}
Marta Bell{\'e}s-Mu{\~n}oz, Miguel Isabel, Jose~Luis Mu{\~n}oz-Tapia, Albert
  Rubio, and Jordi Baylina.
\newblock Circom: A circuit description language for building zero-knowledge
  applications.
\newblock {\em IEEE Transactions on Dependable and Secure Computing},
  20(6):4733--4751, 2022.

\bibitem{chin2021leo}
Collin Chin, Howard Wu, Raymond Chu, Alessandro Coglio, Eric McCarthy, and Eric
  Smith.
\newblock Leo: A programming language for formally verified, zero-knowledge
  applications.
\newblock {\em Cryptology ePrint Archive}, 2021.

\bibitem{ozdemir2022circ}
Alex Ozdemir, Fraser Brown, and Riad~S Wahby.
\newblock Circ: Compiler infrastructure for proof systems, software
  verification, and more.
\newblock In {\em IEEE Symposium on Security and Privacy (SP)}, pages
  2248--2266, 2022.

\bibitem{amin2023lurk}
Nada Amin, John Burnham, Fran{\c{c}}ois Garillot, Rosario Gennaro, Daniel
  Rogozin, Cameron Wong, et~al.
\newblock Lurk: Lambda, the ultimate recursive knowledge.
\newblock {\em Cryptology ePrint Archive}, 2023.

\bibitem{eberhardt2018zokrates}
Jacob Eberhardt and Stefan Tai.
\newblock Zokrates-scalable privacy-preserving off-chain computations.
\newblock In {\em 2018 IEEE International Conference on Internet of Things},
  pages 1084--1091, 2018.

\bibitem{halo2ce}
Privacy \&~Scaling Explorations.
\newblock halo2 community edition.
\newblock Github \url{https://github.com/privacy-scaling-explorations/halo2},
  2023.

\bibitem{zksecurity2023noname}
zksecurity.
\newblock Noname: a programming language to write zkapps.
\newblock \url{https://github.com/zksecurity/noname}, 2023.

\bibitem{kosba2018xjsnark}
Ahmed Kosba, Charalampos Papamanthou, and Elaine Shi.
\newblock xjsnark: A framework for efficient verifiable computation.
\newblock In {\em 2018 IEEE Symposium on Security and Privacy (SP)}, pages
  944--961, 2018.

\bibitem{mina2021o1js}
o1~labs.
\newblock Typescript framework for zk-snarks and zkapps.
\newblock GitHub \url{https://github.com/o1-labs/o1js}, 2021.

\bibitem{scroll2023}
Scroll.
\newblock Scroll zkevm, 2023.
\newblock \url{https://scroll.io/}.

\bibitem{polygon}
polygon.
\newblock Github \url{https://polygon. technology/polygon-zkevm}, 2023.

\bibitem{era2023}
Matter Labs.
\newblock zksync era, 2023.
\newblock \url{https://era.zksync.io/}.

\bibitem{bruestle2023riscZeroZkVM}
Jeremy Bruestle, Paul Gafni, and the RISC Zero~Team.
\newblock Risc zero zkvm: Scalable, transparent arguments of risc-v integrity.
\newblock \url{https://dev.risczero.com/proof-system-in-detail.pdf}, 2023.

\bibitem{arun2024jolt}
Arasu Arun, Srinath Setty, and Justin Thaler.
\newblock Jolt: Snarks for virtual machines via lookups.
\newblock In {\em International Conference on the Theory and Applications of
  Cryptographic Techniques (EUROCRYPT)}, pages 3--33, 2024.

\bibitem{goldberg2021cairo}
Lior Goldberg, Shahar Papini, and Michael Riabzev.
\newblock Cairo--a turing-complete stark-friendly cpu architecture.
\newblock {\em Cryptology ePrint Archive}, 2021.

\bibitem{zhang2023polynomial}
Yuncong Zhang, Shi-Feng Sun, Ren Zhang, and Dawu Gu.
\newblock Polynomial iops for memory consistency checks in zero-knowledge
  virtual machines.
\newblock In {\em International Conference on the Theory and Application of
  Cryptology and Information Security (AsiaCrypt)}, pages 111--141, 2023.

\bibitem{bootle2018arya}
Jonathan Bootle, Andrea Cerulli, Jens Groth, Sune Jakobsen, and Mary Maller.
\newblock Arya: Nearly linear-time zero-knowledge proofs for correct program
  execution.
\newblock In {\em International Conference on the Theory and Application of
  Cryptology and Information Security (AsiaCrypt)}, pages 595--626. Springer,
  2018.

\bibitem{groth2017snarky}
Jens Groth and Mary Maller.
\newblock Snarky signatures: Minimal signatures of knowledge from
  simulation-extractable snarks.
\newblock In {\em Annual International Cryptology Conference}, pages 581--612,
  2017.

\bibitem{gabizon2021fflonk}
Ariel Gabizon and Zachary~J Williamson.
\newblock fflonk: a fast-fourier inspired verifier efficient version of plonk.
\newblock {\em Cryptology ePrint Archive}, 2021.

\bibitem{danezis2014square}
George Danezis, C{\'e}dric Fournet, Jens Groth, and Markulf Kohlweiss.
\newblock Square span programs with applications to succinct nizk arguments.
\newblock In {\em International Conference on the Theory and Application of
  Cryptology and Information Security (AsiaCrypt)}, pages 532--550, 2014.

\bibitem{backes2015adsnark}
Michael Backes, Manuel Barbosa, Dario Fiore, and Raphael~M Reischuk.
\newblock Adsnark: Nearly practical and privacy-preserving proofs on
  authenticated data.
\newblock In {\em IEEE Symposium on Security and Privacy}, pages 271--286,
  2015.

\bibitem{bitansky2013recursive}
Nir Bitansky, Ran Canetti, Alessandro Chiesa, and Eran Tromer.
\newblock Recursive composition and bootstrapping for snarks and proof-carrying
  data.
\newblock In {\em Proceedings of the annual ACM symposium on Theory of
  computing}, pages 111--120, 2013.

\bibitem{circomlib}
circomlib contributors.
\newblock \texttt{circom} compiler, 2022.
\newblock https://github.com/iden3/circomlib.

\end{thebibliography}

\appendix

\section{Sudoku example for ITP}
\label{app: ITP}
\new{
	\mypara{Scenario} When convincing someone that a Sudoku puzzle has a unique solution, we can use IP, PCP or IOP and compare their difference.
	
	\mypara{IP} The verifier can ask any question she likes to the prover who has the complete solution, such as:
	\begin{itemize}[noitemsep, topsep=2pt, partopsep=0pt,leftmargin=0.4cm]
		\item "What's the number in row 3, column 5?"
		\item "Why can't the number 8 be in the 7th box?"
		\item "Explain how you deduced the number in row 2, column 1?"
	\end{itemize}
	
	\mypara{PCP} The prover writes the complete solution on a very large piece of paper (the PCP proof). The verifier is allowed to randomly choose a few cells to check (random oracle access to the proof):
	\begin{itemize}[noitemsep, topsep=2pt, partopsep=0pt,leftmargin=0.4cm]
		\item "Check the number in row 2, column 8."
		\item "Check the number in row 6, column 3."
		\item "Check the number in row 9, column 9."
	\end{itemize}
	
	\mypara{IOP} The prover also provides oracles like PCP but the verifier has more kinds of interactions. 
	\begin{itemize}[noitemsep, topsep=2pt, partopsep=0pt,leftmargin=0.4cm]
		\item First, the prover writes some hints on several sheets of paper (oracles), such as "the sum of each row and column is 45", "each box contains digits from 1 to 9", or a specific deduction step.
		\item Then, the verifier can ask questions about the hints, such as "show me the arrangement of numbers in row 3", or "show me the numbers in box 5".
		\item Last, the verifier can randomly check parts of the hints provided.
	\end{itemize}
}
\section{Library Surveys}
\label{sec:applib}
We survey each library in detail, including \lib{libsnark}~\cite{libsnark}, 
\lib{bellman}~\cite{bellman}, 
\lib{libSTARK}~\cite{libsTark},
\lib{dalek}~\cite{dalek-bulletproofs},
\lib{libiop}~\cite{libiop}, \lib{snarkjs}~\cite{snarkjs}, \lib{gnark}~\cite{gnark}, 
\lib{arkworks}~\cite{arkworks},
\lib{halo2}~\cite{halo2}, 
\lib{Spartan}~\cite{spartan}, 
and \lib{plonky2}~\cite{plonky2}. We discuss the challenges we encountered when implementing the sample programs and elaborate on limitations noted in the tables on the overall usability of each library. We compare the differences between academic and commercial projects and address recommendations to help the developer improve their projects. We also mention the history and the great contributions those projects made to \zk field. 

%We provide the detailed discussion in our open source materials in \url{https://doi.org/10.5281/zenodo.14682405} for interested readers.

% \begin{takeaway}[Recommended Practices]
	%     Across the course of our replication
	%     study, we have pinpointed several concerns in the main steps to construct an application that is universal in all libraries. 
	%     \begin{itemize}[noitemsep, topsep=2pt, partopsep=0pt,leftmargin=0.4cm]
		%         \item \textbf{Find a compiler.} It often remains unclear how to express the statements to be proved as R1CS or plonkish circuits. If no explicit compiler (i.e., DSL) is given, we recommend using the library's built-in gadgets as an alternative.
		%         \item \textbf{Frontend and backend are not separated.} The frontend functions as arithmetizing the circuit constraints to polynomials and field elements, and the backend selects a proving system. In the frontend, the user needs to choose an appropriate curve to avoid data overflow and to be compatible with backend schemes. However, we have found very few documents addressing this problem thus we recommend double configuration before implementation.
		
		%     \end{itemize}
	% \end{takeaway}

\subsection{libsnark}
\lib{libsnark}~\cite{libsnark} is a C++ project started in 2014 and is the first project that aims to provide comprehensive support for {\zk}s. The schemes~\cite{ben2014succinct,groth2016size,danezis2014square,groth2017snarky,backes2015adsnark,bitansky2013recursive} implemented in \lib{libsnark} are all based on QAP because at that time only QAP-based schemes are efficient. \lib{libsnark} offers a great overlook of QAP-based schemes, as it not only includes the most popular Groth16~\cite{groth2016size} scheme but also provides comparisons with former related works.

\mypara{Toolkits} The core toolkit in \lib{libsnark} involves \texttt{relation} (defining the representation of NP language), \texttt{reduction} (containing functions that convert each relation), \texttt{ppzksnark} (implementing the core proof systems), and \texttt{gadget} (simplifying the procedure of specified constraints). \lib{libsnark} defines various relations like \text{R1CS}, \text{TBCS} (Two-input Boolean Circuit Satisfiability), \text{USCS} (Unitary-Square Constraint System), \text{RAM} (Random Access Model), etc. A notable feature of \lib{libsnark} is that those relations can be converted from one to another using functions in \texttt{reduction}. Nowadays, the relations except \text{R1CS} and \text{RAM} are rarely used due to efficiency issues, and our tests are based on \text{R1CS}. \new{ In the compiler's syntax, \lib{libsnark} recommends defining the circuit using its gadgets interface, which allows us focus on the function level of our constraints and not a circuit level.
	In the \texttt{gadget}, \lib{libsnark} implements four useful data types such as variables, array, linear combination variable, linear combination array and arithmetic and loop operations for these types. All the variables are defined in C++ template and need to be instantiated using a specific curve. During our test, we find the syntax is powerful as we define the SHA2 circuit in less than 40 lines.}

\mypara{Documents and Support} The documentation in \lib{libsnark} for installing and testing is abundant. \lib{libsnark} carefully documents all its dependency libraries and discusses whether or not many platforms are compatible. However, there is not enough documentation about how to generate a proof for a new computation. The \texttt{gadget} only contains examples for the inner product and has no comments about how to use other gadget functions. Besides, \lib{libsnark} mentions that it is capable of providing complex high-level language. The language should first be compiled into an R1CS and then linked with \lib{libsnark}, but we find no documentation of this procedure. There are issues opened about this problem, but this library does not have online support. Overall, we solve this issue by reading the source code of the gadget in \texttt{gadget.hpp} and finding more useful gadgets like comparison and multiplex to implement the sample program.

\mypara{Recommendation} We recommend \lib{libsnark} for research or studying \zk. It has an explicit structure of circuits, providing systems and useful tools like gadgets. The subsequent libraries all have a similar structure. \lib{libsnark} is not suitable for implementing proofs for arbitrary applications because it lacks documents about linking an external R1CS. Additionally, we recommended that \lib{libsnark} includes a naming comparison table for \zk schemes, such as mapping \textit{Groth16} to \textit{r1cs\_gg\_ppzksnark}. Currently, this mapping is not documented but only explained in the comments of \textit{r1cs\_gg\_ppzksnark.hpp}, which may confuse users.

\subsection{bellman}
\lib{bellman} \cite{bellman} is a commercial library established in 2017 which implemented groth16~\cite{groth2016size}. \lib{bellman} provides circuit traits and primitive structures, as well as basic gadget implementations such as booleans and number abstractions. \lib{bellman} is currently in its infancy and can only be used to construct simple constraints with its low-level APIs. The goal of \lib{bellman} in 2017 is to pinpoint the future of zcash~\cite{sasson2014zerocash} by implementing the most efficient \zk scheme at that time. \lib{bellman} distinguishes its design from \lib{libsnark}'s gadgetlib as its all variable allocation, assignment, and constraint enforcement are done over the same code path which enables a concise and lightweight program. 

\mypara{Toolkits} 
\new{In the compiler's syntax}, \lib{bellman} provides basic circuit synthesis such as linear combination and multiplication in the elliptic curve and finite field arithmetic. The APIs in \lib{bellman} are low-level and do not offer features that would facilitate ease of use for most users.

\mypara{Documents and Support}
There are few documents about \lib{bellman} as its design intention is for research or proof-of-concept. The developers provide online support for installing but no more others. Concurrently, it is not a easy work to implement complex circuits in \lib{bellman}, but implementing simple circuits is straightforward. 

\mypara{Recommendation}
We recommend \lib{bellman} as a learning opportunity for constructing zk-SNARKs safely and efficiently. Besides, some application-level projects are recommended building on top of the \lib{bellman}.

\subsection{libSTARK}
\lib{libSTARK} \cite{libsTark} is an academic library established in 2018 and has implemented the STARK scheme in C++. Its features include scalability, transparency, and post-quantity security. \lib{libSTARK} provides two examples for testing, i.e., DNA fingerprint blacklist and tinyRAM programs. We managed to install \lib{libSTARK} and run its examples but found that \lib{libSTARK} has not provided any support of implementing our own programs.

\mypara{Toolkit} We delve into the source code to see the gadgets contained in \lib{libSTARK}. \new{In the compiler's syntax, \lib{libSTARK} supports NAND gate, field multiplication, and for-loops. In some of its test code repositories, gadget functions are provided.
	However, the gadgets in \lib{libSTARK} are high-level circuits such as SHA2 and there are no useful gadgets for a user to define her own circuits as the library is mainly for testing its own codes. We conclude that \lib{libSTARK} does not provide a compiler like eDSL as the support of circuit functions is limited.
}

\mypara{Documentation and Support} There are limited documentation on \lib{libSTARK} for user and developer. \lib{libSTARK} has many functionalities, but it is unclear how to use them. Concurrently, \lib{libSTARK} is not under development.

\mypara{Recommendation} We do not recommend \lib{libSTARK} for general use, both for study or real world application. It is because \lib{libSTARK} has not been updated for years and may have serious issues. However, for expert researchers or C++ programmers who want to implement complex circuits using basic gadgets, better to read \lib{libSTARK} source code carefully.

\subsection{dalek}
\lib{dalek}~\cite{dalek-bulletproofs} is a Rust library started in 2018 that aims to have the fastest range proofs with Bulletproofs~\cite{bunz2018bulletproofs}. \lib{Dalek-bulletproof} accelerates range proofs by choosing \texttt{curve25519-dalek} instead of \texttt{libsecp256k1} to enable parallel execution. The APIs provided by \lib{dalek} library are succinct and implementing a new range proof application is straightforward. \lib{dalek} only focuses on range proofs and inner product proofs and does not have R1CS constraints for general circuits.

\mypara{Toolkits} \lib{dalek} provides support for a single range proof, aggregating range proofs and inner product proofs. The range proof aims to demonstrate that an unsigned variable (32 or 64 bits in the \lib{dalek} library) is in a certain range. Aggregated range proofs allow multiple range proofs to be combined into a single proof, and aggregating optimizers can be used to shrink the proof size. The inner product proof aims to demonstrate that a vector $c$ is the inner product of $a$ and $b$, without revealing $b$. \lib{dalek} does not focus on proving constraints for high-level languages.

\mypara{Documents and Support} \lib{dalek}'s documents comprise two parts. The first is telling how range proof and inner product proof work theoretically, and the second is the user-facing documentation. We find it a good insight to combine theoretical formulas with practice in \lib{dalek}. However, its user-facing documents are not enough, and it has not provided examples for its three toolkits, except for a simple single range proof example. Moreover, the theoretical background contains only formulas without fine-grained reasonings or well-designed blogs.

\mypara{Recommendation} We recommend \lib{dalek} as a good start for understanding and practicing range proofs. Although \lib{dalek} is mainly for research purposes, it is convenient to create lightweight applications.

\subsection{libiop}
\lib{libiop}~\cite{libiop} is a C++ library created in 2019 that implemented the three latest IOP-based schemes at that time. 
A feature of \lib{libiop} is that it packs three different proving systems as alternatives for users, including aurora~\cite{ben2019aurora}, fractal~\cite{chiesa2020fractal} and ligero~\cite{ames2017ligero}.
However, \lib{libiop} is mainly for research purposes and has not provided a complete toolchain for creating an R1CS circuit. \new{We follow the library's API and test the efficiency of randomly generated circuits with roughly the same quantity of constraints as the sample programs.}

\mypara{Toolkits} To support three proving systems, \lib{libiop} uses a namespace \textit{iop} which contains \textit{aurora\_iop}, \textit{fractal\_iop}, and \textit{ligero\_iop} as specific protocols. The user-level APIs of the three schemes are the same, which makes them convenient to use. \new{In the compiler's syntax, \lib{libiop} only supports defining the circuit using \textit{generate\_random\_R1CS}. To define a specific circuit, one may have to manually write the constraints, computes the R1CS parameters for them and pass it to \lib{libiop}'s data structures which are time-consuming and not realistic. In our test, we generate R1CS using the output of \lib{libsnark}'s compiler and pass it to \lib{libiop} to make sure our experimental data in \lib{libiop} is based on real circuits.
}

\mypara{Documents and Support} There are few documents about the APIs of \lib{libiop}, and all test examples do not have comments for explanation. Most examples generate circuits randomly rather than implementing specific ones, which makes it challenging to develop practical applications and products. The developers provide online support, but the issues have not been solved for a long period of time.

\mypara{Recommendation} We recommend \lib{libiop} for curious researchers to study the implementation details of such three schemes.

\subsection{snarkjs}
\lib{snarkjs}~\cite{snarkjs} is a javascript library started in 2019 which implements Groth16~\cite{groth2016size}, Plonk~\cite{gabizon2019plonk}, and FFlonk~\cite{gabizon2021fflonk}. The goal of \lib{snarkjs} is to provide comprehensive ZK toolchains for website and blockchain scenarios. In zero-knowledge concepts, snarkjs provides a compiler called \texttt{circom}, and the syntax is similar to C or javascript. 

\mypara{Toolkits} \texttt{Circom} in \lib{snarkjs} is a powerful HDL compiler. It allows users to independently write their constraints in a file with a high-level language (e.g., DSL) that is similar to C, JavaScript and Verilog. When writing \texttt{circom} codes, the user is not required to use snarkjs APIs, which distinguishes \lib{snarkjs} from all other \zk libraries. \new{In the compiler's syntax, all variables are modeled as wire signals. The function in \texttt{circom} is defined as a subcircuit through the keywords \textit{template} and the output is defined by $component$. Some syntax in \texttt{circom} is counter-intuition at the beginning but one can define more complicated constraints once familiar.
}
When using \texttt{circom}, we successfully implement our sample programs and then compile them to R1CS for testing. Other \zk libraries, such as \lib{arkworks}~\cite{arkworks}, are also gradually starting to support \texttt{circom}.

\mypara{Documents and Support} \lib{snarkjs} provides a detailed document about its DSL and its proving systems. It also offers a \texttt{circom} library named circomlib~\cite{circomlib}, which implements many common cryptographic primitives. It has been updated recently and provides timely online support.

\mypara{Recommendation} We recommend using \lib{snarkjs} for proving large and complex computations, as defining a circuit in \texttt{circom} is as convenient as writing a C program. For research or study purposes, \lib{gnark}~\cite{gnark} and \lib{halo2}~\cite{halo2} are better because \lib{snarkjs} does not provide walk-through tutorials, and the basic examples are also not enough.

\subsection{gnark}
\lib{gnark}~\cite{gnark} is a high-performance, open-source Golang library for creating zero-knowledge proofs, particularly \zk applications originating from 2022. \lib{gnark} implements two schemes, Groth16~\cite{groth2016size} and Plonk~\cite{gabizon2019plonk}. One main feature is that \lib{gnark} provides a high-level language for specifying the proof's logic, and its APIs allow developers to easily create, verify, and deploy zero-knowledge proofs. It also includes a built-in compiler that transforms the high-level language into a low-level representation that can be run on various platforms. \lib{gnark} aims to be user-friendly and has various tutorials, including an executable playground for beginners to learn its programming style.

\mypara{Toolkits} \lib{gnark} provides two relevant classes, frontend and backend. The output of the frontend is a preprocessed circuit. With the circuit, the backend chooses a proving system, assigns a valid witness, and outputs a proof. In the frontend, \lib{gnark} provides a DSL specified in \textit{api} class, which is convenient to add constraints. \new{In its syntax, most of the data structures defined in Go can be operated by the \textit{api} interface and recorded in a circuit. With this flexibility, \lib{gnark} claims it has no need for gadgets, because the functions of circuit are implemented in a Go package like any other piece of code.}
Additionally, \lib{gnark} provides a set of pre-built circuit components, such as SHA256 and elliptic curve arithmetic, that can be used to build more complex circuits. In the backend, \lib{gnark} continues to optimize the performance of the two schemes, and we find efficiency improvement in our tests.

\mypara{Documents and Support} The documentation of \lib{gnark} ranges from user's documents to developer's documents. The documentation not only contains guidelines for installing, running, and writing sample codes but also the paradigm of designing \lib{gnark} and implementation details. However, there are some limitations. Firstly, the principle of selection elliptic curve is implicit. \lib{gnark} provides seven curves but has not documented how to choose an appropriate curve to avoid overflow. Secondly, the DSL syntax in \lib{gnark} is implicit. The only way to specify circuits in the DSL is through the \textit{frontend.API} class, but the usage of its internal functions remains unclear. Luckily, \lib{gnark} has full online support, and we find some solutions from its closed issues.

\mypara{Recommendation} We recommend \lib{gnark} for both study, research purposes and commercial use as it has relatively better support from the developers. Concurrently, \lib{gnark} has only implemented two academic schemes, and we believe there will be more in the future.

\subsection{arkworks}
\lib{arkworks}~\cite{arkworks} is a Rust ecosystem for \zk programming that started in 2022. It implements several latest academic \zk approaches including Groth16~\cite{groth2016size}, Plonk~\cite{gabizon2019plonk}, marlin~\cite{chiesa2020marlin}, gm17~\cite{groth2017snarky}, gemini~\cite{bootle2022gemini}, and Bulletproofs~\cite{bunz2018bulletproofs}. The schemes implemented in \lib{arkworks} span various categories and exhibit diverse properties, such as transparency, small proof size, URS, elastic proofs, and post-quantum security.

\mypara{Toolkits} \lib{arkworks} provides an explicit toolchain for compiling circuits and choosing proof systems. In the compiling phase, there are three predefined configures: finite field, elliptic curve, and polynomial, and we document our choice in our project. \new{In the compiler's syntax, \lib{arkworks} supports variable, array, function as data structures and operations like basic arithmetic, for-loop, while-loop and built-in gadgets like inner product proof and hashing.}
When choosing proof systems, \lib{arkworks} provides several sublibraries separating different categories of \zk approaches. 
It also provides a repository binding to \texttt{circom}'s R1CS, facilitating the generation of Groth16 Proof and Witness generation in Rust.

\mypara{Documentation and Support} There are redundant documents, tutorials, and blogs about the algebra, constraint systems, R1CS gadgets, and proof systems of Groth16~\cite{groth2016size}. Developers also continue to solve the issues in each sub-libraries. However, there is still a lack of tutorials and examples for other \zk schemes, such as marlin~\cite{chiesa2020marlin} and gm17~\cite{groth2017snarky}.

\mypara{Recommendation} We recommend \lib{arkworks} for general use. Both for beginners, researchers, and industries as \lib{arkworks} has the best ecosystem in \zk industry. We also recommend \lib{arkworks} to design more circom-compatible for different proof systems. 

\subsection{halo2}
\lib{halo2}~\cite{halo2} is a Rust library developed by Electric Coin Company (\text{ECC}). \lib{halo2} introduces new features and improvements based on Plonk~\cite{gabizon2019plonk}, including recursive proofs and parallel computation. Due to the long review cycle in submission, \lib{halo2} has not been published yet, but a detailed online book is provided to demonstrate its design~\cite{halo2book}.

\mypara{Toolkits} \lib{halo2} provides many available embedded functions for building a circuit. \new{In the compiler's syntax, }it implements redundant gate-level constraints, including addition, multiplication, array, sum, etc. There are also high-level constraints like the inner product and range check (with lookup tables). For popular zero-knowledge applications like hash and signature, \lib{halo2} provides an integrated API as a part of its tools.

\mypara{Documents and Support} All user-level data types and functions in \lib{halo2} are well documented with many executable examples. \lib{halo2} also provides a step-by-step tutorial walk-through and all preliminaries, including Rust tutorials for beginners. The \lib{halo2} handbook is a developer document providing design details in theory and its corresponding source code. The developers also provide online support through email and GitHub issues.

\mypara{Recommendation} 
\lib{halo2} is concurrently a mature, well-documented, and efficient library used in zcash~\cite{sasson2014zerocash}. We recommend \lib{halo2} for both study and commercial purposes. One drawback of \lib{halo2} is that it contains only one proving system, and for users, there are no alternatives.

\subsection{Spartan}
\lib{Spartan}~\cite{spartan} is a Rust library from Microsoft originating in 2020 that aims to implement transparent {\zk}s with a fast prover. \lib{Spartan} supports generating R1CS randomly for testing and also provides a compiling toolchain from high-level programs of interest. 

\mypara{Toolkits} \lib{Spartan} is claimed to be general, high performance, and with a complete toolchain of compiling high-level programs. \new{In the compiler's syntax, the data structures and operations are similar to \lib{halo2} \cite{halo2} as they both utilize features of Rust. Compared to \lib{halo2}, \lib{Spartan} provides less gadget functions as it mainly focuses on optimizing the efficiency of proving systems for R1CS.}
We test its efficiency by a built-in random circuit generator. However, there are no tutorials about its related language syntax and how to use the compiler. The default example defining a simple circuit is very complex compared to other libraries.

\mypara{Documents and Support} There is little documentation about the programming pattern of \lib{Spartan}. The inputs and outputs of prover transcripts remain unclear, and we find many users struggle to compile R1CS for their own applications. A popular hash application in its documentation, however, is not provided as an example. 

\mypara{Recommendation} \lib{Spartan} achieves high performance but is not capable of supporting commercial use since this library has not received a security review or audit. We recommend users focus on the syntax of \lib{Spartan}'s DSL, which wish to compile a R1CS on their own. 

\subsection{plonky2}
\lib{plonky2}~\cite{plonky2} is a Rust library that started in 2023, where the Polygon team implemented several optimizations for the Plonk scheme. The goal of \lib{plonky2} is to scale zero-knowledge proof in Ethereum layer 2 rollup by providing fast, efficient, and secure proving systems. The main feature of Plonky is that it utilizes a recursive proof technique. Recursive proof is a space-saving technique, and it allows the prover to demonstrate many statements at once without increasing the proof size, which is indeed the requirement in zk-rollup. 

\mypara{Toolkits} \new{In the compiler's syntax, \lib{plonky2} contains the data structures and operations similar to \lib{halo2} \cite{halo2} and \lib{Spartan} \cite{spartan} and we find the same programming pattern when defining circuits using the Rust-based eDSLs.} To utilize the recursive feature, \lib{plonky2} defines an extra class \textit{cyclic\_circuit} for the proof and verification.

\mypara{Documents and Support} \lib{plonky2} is a commercial project for Ethereum developers and is still in development. The documents now only elaborate on its features but without API details. The only way for independent developers to compile \lib{plonky2} is to walk through online examples. However, the recursive proof example is incomplete and cannot be extended to generic circuits.

\mypara{Recommendation} We only recommend \lib{plonky2} for developing Ethereum ZK rollup applications as, at this time, the only way to use Plonky API is by reading its source code.

%\input{1.intro}
%\input{2.background}
%\input{3.overview}
%\input{4.technique}
%\input{5.implementation}  
%\input{6.discussion}
%\input{8.ethicalState}

%\input{7.appendix}
%\input{9.libraries}
%%%%%%%%%%%%%%%%%%%%%%%%%%%%%%%%%%%%%%%%%%%%%%%%%%%%%%%%%%%%%%%%%%%%%%%%%%%%%%%%
\end{document}